\documentclass[sigconf]{acmart}
\usepackage{mathtools}
\usepackage{amsthm}
\usepackage{multirow}   
\usepackage{subcaption}
\usepackage{pifont}
\usepackage{colortbl}
\usepackage{xcolor}

\definecolor{bg_ref}{HTML}{F7F7F7}    
\definecolor{bg_bad}{HTML}{FFF5E6}    
\definecolor{bg_sota}{HTML}{F0F6FF}   
\definecolor{bg_ours}{HTML}{E8FDF5}   
\definecolor{bgGreen}{HTML}{D9EAD3} 
\definecolor{bgRed}{HTML}{F4CCCC}   
\definecolor{bgBlue}{HTML}{E8F0FE}  
\newcommand{\cmark}{\ding{51}}
\newcommand{\xmark}{\ding{55}}

\copyrightyear{2026}
\acmYear{2026}
\setcopyright{cc}
\setcctype{by}
\acmConference[MM '26]{Proceedings of the 34th ACM International Conference on Multimedia}{November 10--14, 2026}{Rio de Janeiro, Brazil}
\acmBooktitle{Proceedings of the 34th ACM International Conference on Multimedia (MM '26), November 10--14, 2026, Rio de Janeiro, Brazil}
\acmDOI{10.1145/3767308.3836568}
\acmISBN{979-8-4007-2213-4/2026/11}

\author{Hao Ma}
\affiliation{
  \institution{Institute of Artificial Intelligence, China Telecom (TeleAI)}
  \city{Shanghai}
  \country{China}
}
\email{mahao2000818@hotmail.com}

\author{Ruihao Jing}
\affiliation{
  \institution{Institute of Artificial Intelligence, China Telecom (TeleAI)}
  \city{Shanghai}
  \country{China}
}
\email{ruihaojing@mail.nwpu.edu.cn}

\author{Shansong Liu}
\affiliation{
  \institution{Institute of Artificial Intelligence, China Telecom (TeleAI)}
  \city{Shanghai}
  \country{China}
}
\email{dadinghh2@gmail.com}

\author{Cheng Gong}
\affiliation{
  \institution{Institute of Artificial Intelligence, China Telecom (TeleAI)}
  \city{Shanghai}
  \country{China}
}
\email{gongc6@chinatelecom.cn}

\author{Chi Zhang}
\affiliation{
  \institution{Institute of Artificial Intelligence, China Telecom (TeleAI)}
  \city{Shanghai}
  \country{China}
}
\email{zhangc120@chinatelecom.cn}

\author{Xiao-Lei Zhang}
\authornote{Corresponding authors.}
\affiliation{
  \institution{Institute of Artificial Intelligence, China Telecom (TeleAI)}
  \city{Shanghai}
  \country{China}
}
\email{xiaolei.zhang@nwpu.edu.cn}

\author{Xuelong Li}
\authornotemark[1]
\affiliation{
  \institution{Institute of Artificial Intelligence, China Telecom (TeleAI)}
  \city{Shanghai}
  \country{China}
}
\email{xuelong\_li@ieee.org}

\renewcommand{\shortauthors}{Hao Ma et al.}
\begin{abstract}
High-fidelity general audio coding at ultra-low bitrates is crucial for applications ranging from low-bandwidth communication to audio-language modeling. However, existing neural audio codecs struggle under extreme bitrate constraints, exhibiting degraded performance: weak semantic alignment leads to semantic distortion, while limited generative capacity compromises acoustic fidelity. To address these challenges, we propose TeleSound‑Codec, a scalable general audio codec that preserves semantic integrity and fine acoustic details together via language‑grounded supervision and scalable generative modeling. Specifically, to achieve strong semantic alignment, we propose to train a semantic encoder under the guidance of a pre-trained large language model. This language-grounded supervision grounds codec latents in human-interpretable linguistic spaces, enhancing semantic integrity and facilitating downstream applications. To improve generative capacity, we further propose training a flow-matching-based generative decoder jointly with an acoustic encoder that captures residual acoustic details. This scalable generative modeling design enables reconstruction quality to scale with computation. Experiments demonstrate that, when scaling the model size to 1.8B parameters, TeleSound-Codec achieves high-fidelity reconstruction of 32kHz general audio—including speech, music, and other sound types—at bitrates as low as 0.275kbps. We present demos on: https://aisaka0v0.github.io/codec\_demo/.
\end{abstract}

\begin{document}
\title[High-Fidelity General Audio Coding Under 1kbps via Language-Grounded Supervision and Scalable Generative Modeling]{TeleSound-Codec: High-Fidelity General Audio Coding Under 1kbps via Language-Grounded Supervision and Scalable Generative Modeling}
\maketitle
\begin{figure}
    \centering
    \includegraphics[width=0.75\linewidth]{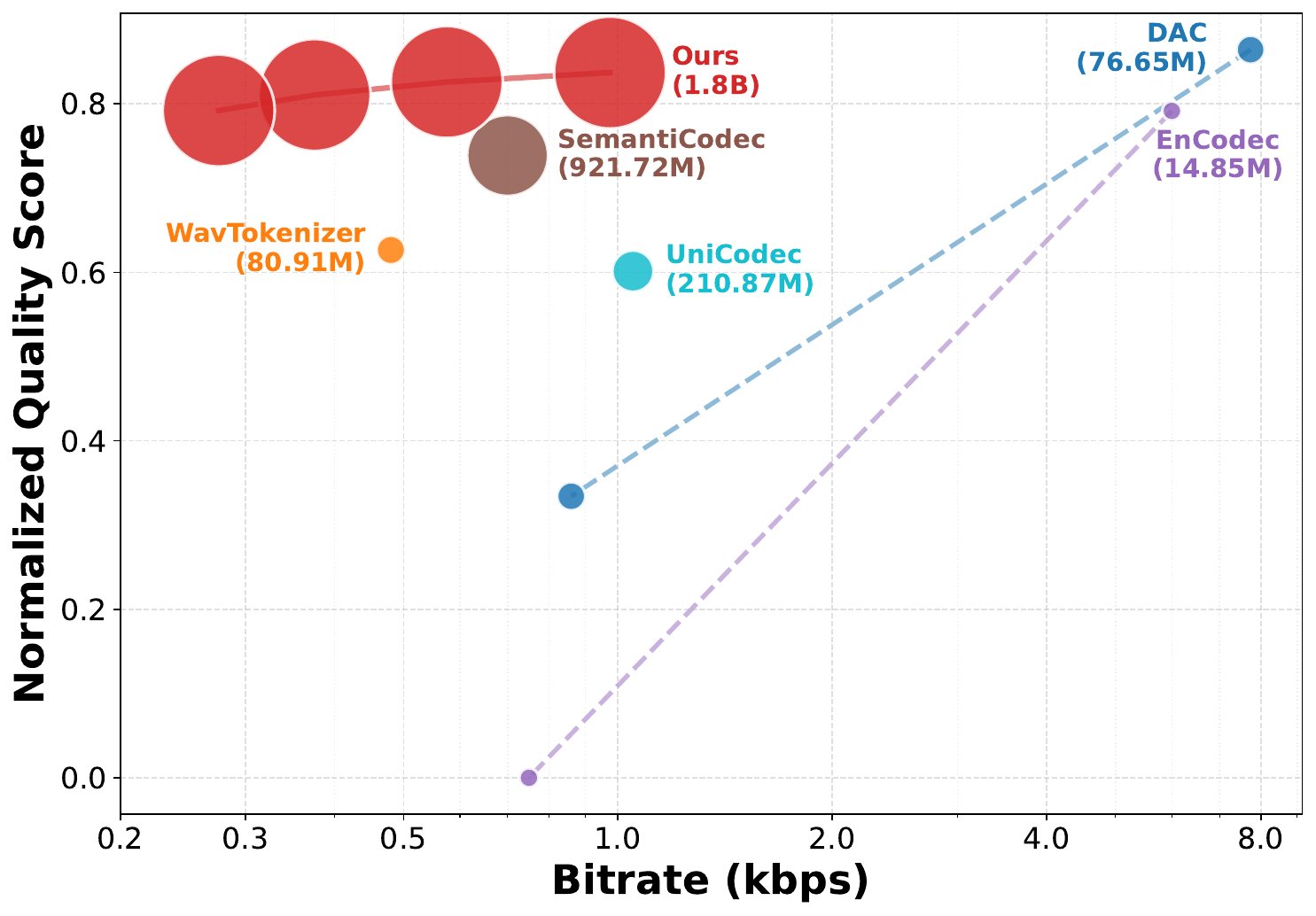}
    \vspace{-4mm}
    \caption{Rate-quality comparison of various audio codecs. The y-axis denotes the average normalized objective metrics across \textit{speech}, \textit{sound}, and \textit{music} domains. Marker size reflects the number of parameters. Our method consistently outperforms baselines, achieving high quality in \emph{sub-1kbps}.}
    \Description{A plot comparing rate versus quality for multiple audio codecs. The proposed method achieves the best quality at very low bitrates, below 1 kbps, across speech, sound, and music domains. Larger markers indicate models with more parameters.}
    \label{fig:0}
    \vspace{-4mm}
\end{figure}
\section{Introduction}

Audio coding aims to represent continuous audio signals in a compact form while preserving perceptual fidelity. Traditional audio codecs, such as MP3~\cite{MP3} and AAC~\cite{AAC}, rely on hand-crafted signal processing pipelines and have achieved remarkable success in waveform compression and transmission for decades. These codecs are primarily optimized for reconstruction quality, typically operating at medium to high bitrates of hundreds of kbps in practical settings.
However, emerging applications increasingly demand much more compact audio representations, particularly in scenarios of low-bandwidth communication~\cite{aiflow1,aiflow2}, downstream generative~\cite{borsos2023audiolm, du2024cosyvoice, agostinelli2023musiclm, aiflow_generative1}, and understanding~\cite{zeng2024glm, kimiaudio, aiflow_understanding1, aiflow_understanding2} tasks. In such regimes, conventional signal-processing-based codecs become inefficient and suffer severe semantic and acoustic distortion, as they neither explicitly model high-level semantic information nor maintain acoustic details under aggressive compression.

Recent advances in deep neural networks have introduced a new paradigm for audio coding. Neural audio codecs~\cite{zeghidour2021soundstream} adopt end-to-end learning to automatically discover compact representations, enabling significantly lower bitrates than traditional codecs while offering improved flexibility. These approaches have positioned neural codecs as a promising direction for ultra-low bitrate audio coding beyond waveform-centric compression.
Existing neural audio codecs generally fall into two categories—\emph{acoustic codecs}~\cite{zeghidour2021soundstream, encodec, dac, hificodec} and \emph{semantic codecs}~\cite{zhang2023speechtokenizer, xcodec, chen2025sac, semanti, yang2025almtokenizer}—based on the nature of the information they are designed to preserve. Acoustic codecs aim to reproduce waveform‑level details with high perceptual fidelity, typically employing waveform-centric objective designs to minimize signal distortion. In contrast, semantic audio codecs focus on encoding high-level information invariant to acoustic variations, typically leveraging discrete representations produced by self‑supervised learning (SSL) models~\cite{hsu2021hubert} to achieve higher compression efficiency.

However, achieving high-fidelity general audio coding at ultra-low bitrates remains an open challenge. Existing approaches fail to maintain semantic and acoustic fidelity together, as waveform-centric methods require high token rates for acoustic fidelity, while semantic codecs lose fine-grained acoustic details, leading to perceptual quality loss.
The above weakness was caused by at least the following two technical limitations: 1) \emph{Weak semantic alignment:} most semantic codecs define their semantic representations within an audio-only SSL latent space without explicit alignment to the human-interpretable linguistic space, thereby hindering intelligibility and limiting integration with modern language models and multimodal systems; 2) \emph{Limited generative capacity:} existing codecs typically rely on GAN decoders. They have limited modeling capacity and are difficult to scale to very large models~\cite{dhariwal2021diffusion}. Eventually, they struggle to faithfully recover fine-grained acoustic details from highly abstract representations.

To preserve semantic integrity and fine-grained acoustic details together at ultra-low bitrate constraints, we propose \textbf{TeleSound-Codec}, a novel neural audio codec that features the following two contributions: \emph{language-grounded supervision} and \emph{scalable generative modeling}.
Specifically, we introduce language-grounded supervision by guiding a semantic encoder with a pretrained large language model (LLM), which explicitly grounds codec latents in human-interpretable linguistic spaces and improves semantic integrity under aggressive compression.
To further recover faithful acoustic details, we propose scalable generative modeling consisting of an acoustic encoder and a diffusion Transformer (DiT) based generative decoder~\cite{peebles2023scalable}. The acoustic encoder captures fine-grained residual information that cannot be represented by semantic tokens alone. The generative decoder is trained with a rectified flow matching objective~\cite{liu2022flow} and jointly leverages semantic and acoustic representations for high-fidelity audio reconstruction. After scaling up the model size of the generative decoder, we empirically observe that higher computational and storage capacity can be traded for improving reconstruction quality, enabling state-of-the-art high-fidelity audio reconstruction in the \emph{sub-1kbps} regime, as illustrated in Figure~\ref{fig:0}.
To summarize, our novel contributions include:
\begin{itemize}
  \item We integrate a large language model into the training of a general audio codec.
  \item We reveal a scaling trend of increasing computation for reduced audio transmission bandwidth.
  \item We achieve high-fidelity reconstruction of \textbf{32kHz} general audio at \textbf{0.275kbps}.
\end{itemize}

\section{Related Work}
\subsection{Acoustic Audio Coding}
Neural audio coding compresses continuous audio signals into discrete representations while preserving perceptual quality. Following the VQ-VAE~\cite{vqvae} framework, residual vector quantization (RVQ)~\cite{zeghidour2021soundstream} has become a standard design in neural audio codecs. SoundStream~\cite{zeghidour2021soundstream} and EnCodec~\cite{encodec} established RVQ-based GAN codecs as strong baselines, while later works such as HiFi-Codec~\cite{hificodec} and DAC~\cite{dac} further improved reconstruction fidelity through more efficient quantization and discriminator designs. 
However, these methods primarily optimize waveform-level reconstruction and provide limited semantic structure, which becomes increasingly fragile under aggressive bitrate reduction.

\subsection{Semantic Audio Coding}
Recent studies introduce semantic supervision to enhance the robustness of discrete audio tokens. SpeechTokenizer~\cite{zhang2023speechtokenizer} aligns early RVQ stages with HuBERT~\cite{hsu2021hubert} features to encourage semantic-acoustic disentanglement, while X-Codec~\cite{xcodec} and SAC~\cite{chen2025sac} incorporate pre-trained semantic representations through feature fusion and dual-stream architectures. 
Although these approaches improve intelligibility, they mainly rely on speech-oriented SSL teachers and do not explicitly align latent representations with language-level semantic spaces. On the other hand, LLM-Codec~\cite{yang2024uniaudio} directly quantizes audio features into the discrete sub-word embedding space of a language model, but focuses primarily on probing capabilities rather than reconstruction. Much more recent works, such as MiMo-Audio-Tokenizer~\cite{mimo} and Moss-Audio-Tokenizer~\cite{moss}, leverage text-related proxy tasks to extract semantically rich audio tokens. However, these explorations are still limited to the GAN-based decoders, leading to suboptimal performance in ultra-low-bitrate general audio coding. In contrast, our approach leverages understanding-level supervision to learn semantic representations aligned with linguistic information for general audio and introduces scalable generative modeling for high-fidelity audio reconstruction.
\begin{figure*}
    \centering
    \includegraphics[width=0.7\linewidth]{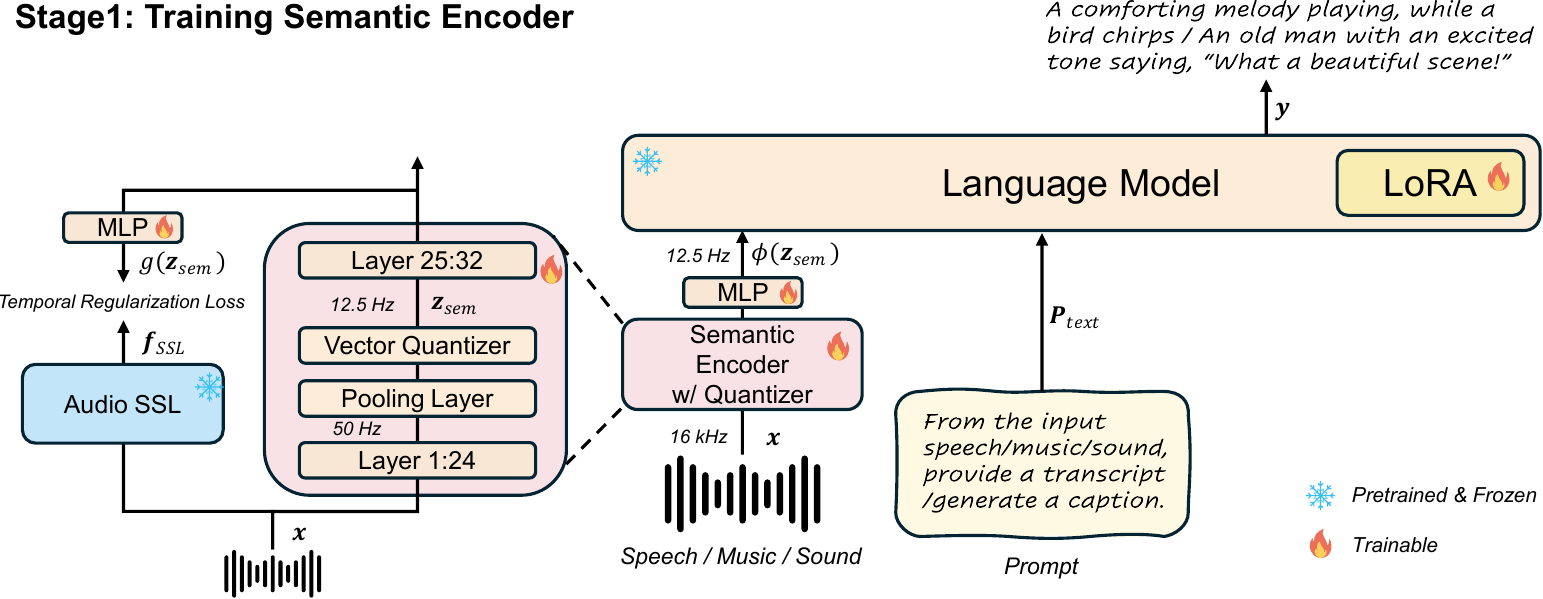}
    \vspace{-2mm}
    \caption{Illustration of the training process of the semantic encoder.}
    \Description{Illustration of the training process of the semantic encoder.}
    \label{fig:main1}
    \vspace{-2mm}
\end{figure*}
\subsection{General Audio Coding at Low Bitrates}
Achieving high-fidelity general audio coding at ultra-low bitrates remains a challenging open problem. While previous methods like WavTokenizer~\cite{ji2024wavtokenizer} and UniCodec~\cite{unicodec} achieve extreme compression through single-quantizer designs, their lack of strong semantic grounding and reliance on conventional GAN decoders often result in degraded perceptual quality and semantic ambiguity. More recently, several works have explored the use of more powerful generative decoders, such as flow matching, to improve acoustic reconstruction quality. Representative examples include S3-Tokenizer~\cite{du2024cosyvoice} and GLM-4-Voice-Tokenizer~\cite{zeng2024glm} in the speech domain, as well as MuCodec~\cite{mucodec} in the music domain. However, these approaches typically rely on purely semantic tokens, overlooking the explicit modeling of fine-grained acoustic details. Moreover, their designs are often tailored to specific audio domains, limiting their ability to generalize to general audio scenarios.
In contrast, SemantiCodec~\cite{semanti} mitigates this by clustering AudioMAE~\cite{audiomae} features and employing diffusion-based decoding for general audio. However, its semantic supervision is limited to audio-only SSL representations. Moreover, how the scalability of its diffusion decoder reflects the minimum bandwidth requirement under certain quality constraints remains underexplored.
These limitations reveal the need for investigating stronger semantic supervision to preserve content integrity, as well as scalable generative modeling for high-fidelity audio reconstruction.

\section{Method}

\begin{table}[t] \centering \caption{Audio understanding tasks and tuning formats.} \label{tab:audio_tasks} \tiny \vspace{-2mm} \begin{tabular}{lp{3.5cm}p{3.2cm}} \toprule \textbf{Domain} & \textbf{Instruct Input} & \textbf{Model Response} \\ \midrule Speech & From the input speech, identify speaker traits and provide a transcript. & A \{{\it age}\} \{{\it gender}\}, speaking in \{{\it language}\}, has a \{{\it high/moderate/low}\} pitched voice and a \{{\it fast/moderate/slow}\} speech rate, expressing a \{{\it emotion}\} tone, saying \{{\it transcription}\}. \\ \midrule Sound & From the input sound, generate an audio caption. & \{{\it audio caption}\} \\ \midrule Music & From the input music, detect genre, moods, instruments, and give a caption. & Genre: \{{\it genre}\}; moods: \{{\it moods}\}; instruments: \{{\it instruments}\}; caption: \{{\it music caption}\}. \\ \bottomrule \end{tabular} \vspace{-2mm} \end{table}

\begin{figure}
    \centering
    \includegraphics[width=0.9\linewidth]{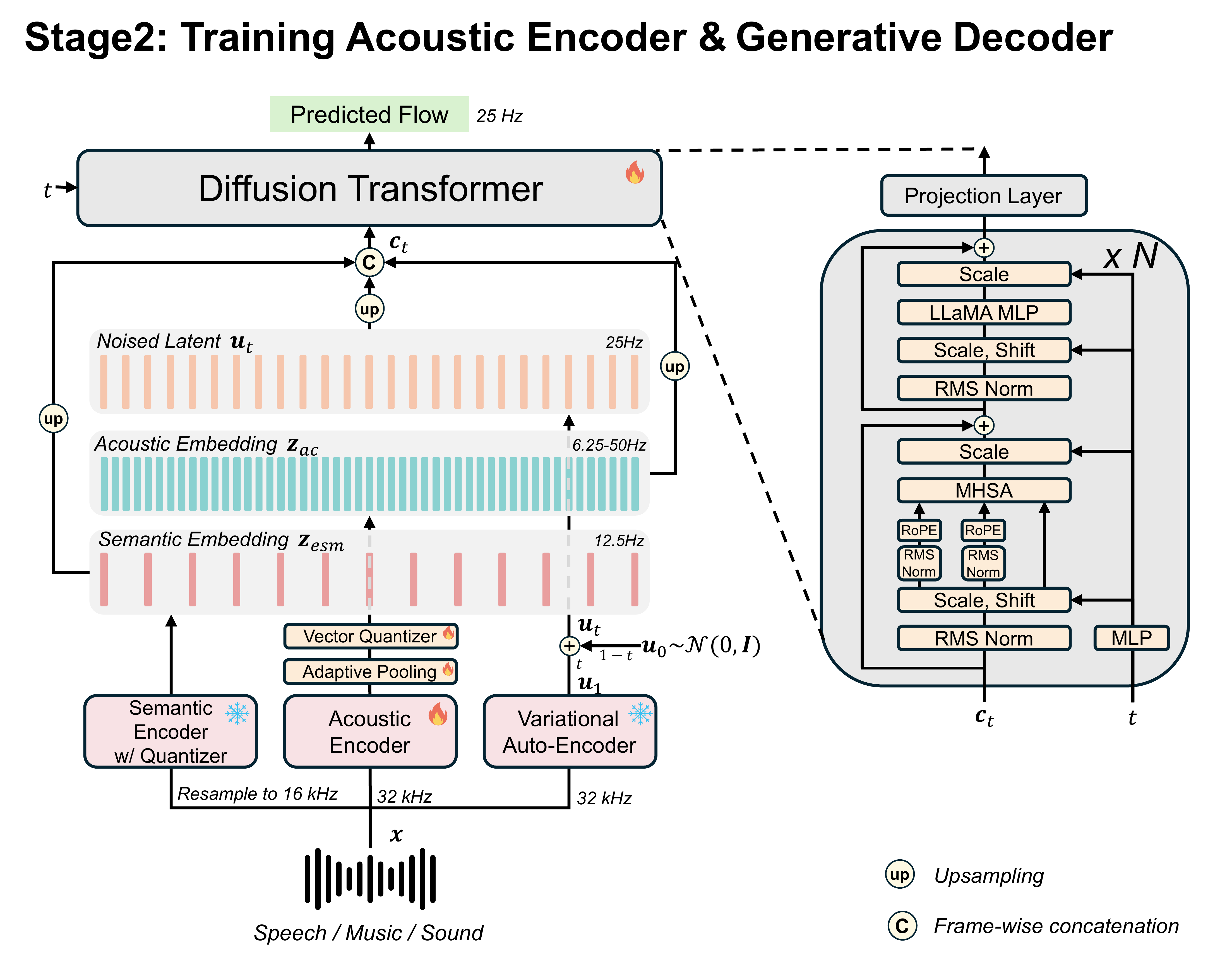}
    \vspace{-2mm}
    \caption{Illustration of the training process of the acoustic encoder along with the generative decoder.}
    \Description{Illustration of the training process of the acoustic encoder along with the generative decoder.}
    \label{fig:main2}
    \vspace{-4mm}
\end{figure}

In this section, we introduce TeleSound-Codec, a two-stage neural audio codec designed for ultra-low-bitrate general audio coding. As illustrated in Figure~\ref{fig:main1} and Figure~\ref{fig:main2}, the proposed framework consists of two phases: 1) \emph{Stage 1}, a semantic-first tokenization stage that extracts discrete, language-grounded semantic tokens under aggressive compression; and 2) \emph{Stage 2}, an acoustic reconstruction stage that models complementary fine-grained acoustic details and synthesizes high-fidelity waveforms via flow matching.

\subsection{Stage 1: Language-Grounded Semantic Tokenization}
\label{sec:stage1}

As shown in Figure~\ref{fig:main1}, the objective of Stage 1 is to obtain an extremely compact discrete representation that both preserves high‑level semantic content and aligns explicitly with a human-interpretable linguistic space spanning speech, music, and general sounds. To this end, we explicitly ground these semantic tokens in a linguistic space using supervision from a pre-trained large language model.

\paragraph{Semantic Encoding and Discretization.}
Given an input waveform $\mathbf{x}$, a semantic encoder extracts high-level representations that are discretized into semantic tokens:
\begin{equation}
\mathbf{z}_{\text{sem}} = Q_{\text{sem}}\big(E_{\text{sem}}(\mathbf{x})\big),
\end{equation}
where $E_{\text{sem}}(\cdot)$ denotes the semantic encoder and $Q_{\text{sem}}(\cdot)$ is a vector quantization operator following SimVQ~\cite{simvq}. These semantic tokens are optimized to capture high-level semantic information while discarding fine-grained acoustic details.

\paragraph{Language-Grounded Supervision.}
To explicitly ground the semantic tokens in the human-interpretable linguistic space, we adopt a pre-trained LLM as a semantic supervisor. The quantized semantic tokens $\mathbf{z}_{\text{sem}}$ are projected into the LLM embedding space via a learnable adaptation layer and treated as soft prompts. These audio-derived prompts are prepended to a task-specific textual prompt $\mathbf{p}_{\text{text}}$ and fed into the LLM:
\begin{equation}
\mathcal{L}_{\text{LM}}
= - \mathbb{E}_{(\mathbf{x}, \mathbf{y})}
\log p_{\text{LLM}}\!\left(
\mathbf{y} \mid [\phi(\mathbf{z}_{\text{sem}}), \mathbf{p}_{\text{text}}]
\right),
\label{eq:lm}
\end{equation}
where $\phi(\cdot)$ projects semantic tokens into the LLM embedding space and $\mathbf{y}$ denotes the corresponding textual description. The model is trained to autoregressively generate a target text sequence $\mathbf{y}$ as detailed in Table~\ref{tab:audio_tasks}. This objective enforces semantic consistency even under aggressive temporal downsampling and quantization.

\paragraph{Temporal Regularization.}
In our preliminary experiments, we observe that relying solely on language supervision often leads to semantic token collapse, especially for non-speech or weakly annotated audio. The textual descriptions for these types of audio are generally sparse and temporally coarse. To preserve temporal structure, we introduce an auxiliary alignment regularization based on a frozen self-supervised audio representation model.

Let $\mathbf{f}_{\text{SSL}}$ denote frame-level features extracted from the audio SSL model. We encourage alignment between semantic tokens and these temporally rich representations using a cosine-distance objective:
\begin{equation}
    \mathcal{L}_{\text{temp}} =
    \frac{1}{T}\sum_{i=1}^{T}
    \left(
    1 - \cos \big(
    g(\mathbf{z}_{\text{sem}}^{(i)}),
    \mathbf{f}_{\text{SSL}}^{(i)}
    \big)
    \right),
\end{equation}
where $g(\cdot)$ is a lightweight projection network together with the upper encoder blocks, which maps semantic tokens to the SSL feature space. This dense frame-level regularization complements the sparse language objective and stabilizes codebook utilization.

\paragraph{Training Objective in Stage 1.}
The semantic encoder $E_{\text{sem}}(\cdot)$, the semantic vector quantizer $Q_{\text{sem}}(\cdot)$, and the adaptation layers $g(\cdot)$, $\phi(\cdot)$ are jointly optimized. The backbone parameters of the pre-trained LLM and the self-supervised audio model are kept frozen, with only the LoRA~\cite{hu2022lora} parameters updated for language supervision.

The overall training objective in Stage 1 is formulated as a weighted combination of the semantic quantization loss, the language modeling loss, and the temporal regularization:
\begin{equation}
    \mathcal{L}_{\text{stage1}} =
    \mathcal{L}_{\text{LM}}
    + \lambda_{\text{VQ}}^{\text{sem}} \mathcal{L}_{\text{VQ}}^{\text{sem}}
    + \lambda_{\text{temp}} \mathcal{L}_{\text{temp}},
\end{equation}
where $\lambda_{\text{VQ}}^{\text{sem}}$ and $\lambda_{\text{temp}}$ control the contributions of the quantization module and the temporal regularization. This multi-objective training encourages the semantic tokens to be \emph{compact}, \emph{language-grounded}, and \emph{temporally coherent} under extreme bitrate constraints. 

\subsection{Stage 2: Acoustic Tokenization and Scalable Generative Reconstruction}
\label{sec:stage2}

As shown in Figure~\ref{fig:main2}, Stage 2 focuses on reconstructing high-fidelity audio by modeling residual acoustic details and performing conditional generative synthesis. Unlike conventional codecs that operate at a fixed acoustic token rate, Stage 2 introduces \emph{adaptive acoustic pooling} to enable flexible bitrate control, and employs \emph{temporally aligned conditional generation} based on a scalable rectified flow DiT to ensure stable and coherent reconstruction.

\paragraph{Adaptive Acoustic Tokenization.}
Given the input waveform $\mathbf{x}$, an acoustic encoder extracts fine-grained acoustic representations:
\begin{equation}
\mathbf{h}_{\text{ac}} = E_{\text{ac}}(\mathbf{x}).
\end{equation}
To support variable bitrates in a unified model, we introduce an adaptive pooling operator that compresses acoustic features with a controllable temporal resolution:
\begin{equation}
\tilde{\mathbf{h}}_{\text{ac}} = \mathcal{P}_s(\mathbf{h}_{\text{ac}}),
\end{equation}
where $\mathcal{P}_s(\cdot)$ denotes pooling with stride $s$. By varying 
$s$, the acoustic token rate can be dynamically adjusted without retraining the model, enabling the reconstruction quality to scale with bitrate. The pooled features are then discretized into acoustic tokens:
\begin{equation}
\mathbf{z}_{\text{ac}} = Q_{\text{ac}}(\tilde{\mathbf{h}}_{\text{ac}}).
\end{equation}
Unlike RVQ, which relies on multiple codebooks, our approach uses a single shared codebook to represent acoustic tokens at different temporal resolutions. This design enables variable-bitrate tokenization without increasing the vocabulary size when integrated with a downstream language model, thereby reducing the embedding overhead when large codebooks are used in the LM-based generation.

\paragraph{Temporally Aligned Conditional Generative Modeling.}
Stage 2 models the conditional distribution of clean acoustic latents given semantic and acoustic tokens:
\begin{equation}
p(\mathbf{u} \mid \mathbf{z}_{\text{sem}}, \mathbf{z}_{\text{ac}}),
\end{equation}
where $\mathbf{u}$ denotes latent representations in a pre-trained variational auto-encoder (VAE)~\cite{kingma2013auto} latent space. We adopt rectified flow matching~\cite{liu2022flow} to parameterize this distribution.

\begin{table*}[t]
  \caption{Objective evaluation results in the speech domain. Evaluations are conducted on SeedTTS-ZH and SeedTTS-EN. SR denotes the audio sampling rate (Hz) that models operate on; TR and BR denote the discrete token rate (Hz) and bitrate (kbps), respectively; Unified indicates whether the model supports general audio coding beyond speech. \emph{Note:} The ``Ours VAE'' serves as the reconstruction upper bound reference, as our codec operates directly within its latent space.}  \label{tab:seedtts_results}
  \vspace{-2mm}
  \centering
  \footnotesize
    \begin{tabular}{lcccccccccccc}
      \toprule
      \multirow{3}{*}{\bf Method}
        & \multirow{3}{*}{\bf SR}
        & \multirow{3}{*}{\bf TR}
        & \multirow{3}{*}{\bf BR}
        & \multirow{3}{*}{\bf Unified}
        & \multicolumn{4}{c}{\bf SeedTTS-ZH}
        & \multicolumn{4}{c}{\bf SeedTTS-EN} \\
      \cmidrule(lr){6-9} \cmidrule(lr){10-13}
        & & & &
        & \multicolumn{2}{c}{\bf MOS$\uparrow$} & \multirow{2}{*}{\bf SIM$\uparrow$} & \multirow{2}{*}{\bf CER (\%)$\downarrow$}
        & \multicolumn{2}{c}{\bf MOS$\uparrow$} & \multirow{2}{*}{\bf SIM$\uparrow$} & \multirow{2}{*}{\bf WER (\%)$\downarrow$} \\
      \cmidrule(lr){6-7} \cmidrule(lr){10-11}
        & & & &
        & {\bf DNS} & {\bf NISQA} & & 
        & {\bf DNS} & {\bf NISQA} & & \\
      \midrule

      \rowcolor{bg_ref}
      \textcolor{gray}{Ground Truth}
        & \textcolor{gray}{-} & \textcolor{gray}{-} & \textcolor{gray}{-} & \textcolor{gray}{-}
        & \textcolor{gray}{3.849} & \textcolor{gray}{4.045} & \textcolor{gray}{1.000} & \textcolor{gray}{5.10}
        & \textcolor{gray}{3.830} & \textcolor{gray}{3.778} & \textcolor{gray}{1.000} & \textcolor{gray}{1.93} \\
      
      \midrule

      Encodec~\cite{encodec}
        & 24k & 600 & 6.000 & \cmark
        & 3.614 & 3.655 & 0.942 & 6.33
        & 3.645 & 3.627 & 0.937 & 2.49 \\
      DAC~\cite{dac}
        & 44k & 774 & 7.740 & \cmark
        & 3.825 & 4.085 & 0.975 & 5.47
        & 3.869 & 3.973 & 0.968 & 2.11 \\
      SpeechTokenizer~\cite{zhang2023speechtokenizer}
        & 16k & 400 & 4.000 & \xmark
        & 3.782 & 3.849 & 0.905 & 7.74
        & 3.845 & 3.925 & 0.897 & 3.17 \\
      
      \midrule
      \multicolumn{13}{c}{\textit{Codecs operating at low bitrates}} \\
      \midrule
      \rowcolor{bg_bad}
      Encodec-RVQ1~\cite{encodec}
        & 24k & 75 & 0.750 & \cmark
        & 2.700 & 1.218 & 0.641 & 218.83
        & 2.775 & 1.134 & 0.597 & 74.86 \\
      \rowcolor{bg_bad}
      DAC-RVQ1~\cite{dac}
        & 44k & 86 & 0.860 & \cmark
        & 3.091 & 2.196 & 0.699 & 167.11
        & 3.248 & 2.371 & 0.668 & 39.63 \\
      \rowcolor{bg_bad}
      SpeechTokenizer-RVQ1~\cite{zhang2023speechtokenizer}
        & 16k & 50 & 0.500 & \xmark
        & 2.984 & 1.023 & 0.548 & 584.07
        & 3.002 & 0.915 & 0.593 & 11.65 \\

      \midrule

      \rowcolor{bg_sota}
      SemantiCodec~\cite{semanti}
        & 16k & 50 & 0.700 & \cmark
        & 3.845 & 3.902 & 0.853 & 14.40
        & 3.869 & 3.838 & 0.818 & 5.93 \\
      \rowcolor{bg_sota}
      WavTokenizer~\cite{ji2024wavtokenizer}
        & 24k & 40 & 0.480 & \cmark
        & 3.757 & 3.615 & 0.761 & 34.98
        & 3.792 & 3.704 & 0.761 & 12.78 \\
      \rowcolor{bg_sota}
      UniCodec~\cite{unicodec}
        & 24k & 75 & 1.050 & \cmark
        & 3.767 & 3.973 & 0.826 & 13.78
        & 3.869 & 4.177 & \textbf{0.830} & 4.80 \\
    \rowcolor{bg_sota}
    MOSS-Audio-Tokenizer~\cite{moss}
        & 24k & 37.5 & 0.375 & \cmark
        & 3.471 & 3.537 & 0.854 & 14.78
        & 3.414 & 3.671 & 0.761 & 6.48 \\
      \midrule
    \rowcolor{bg_ref}
      \textcolor{gray}{Ours VAE}
        & \textcolor{gray}{32k} & \textcolor{gray}{-} & \textcolor{gray}{-} & \textcolor{gray}{\cmark}
        & \textcolor{gray}{3.842} & \textcolor{gray}{4.001} & \textcolor{gray}{0.982} & \textcolor{gray}{6.37}
        & \textcolor{gray}{3.842} & \textcolor{gray}{3.822} & \textcolor{gray}{0.972} & \textcolor{gray}{2.06} \\

      \rowcolor{bg_ours}
      \textbf{Ours} (62.5Hz)
        & 32k & 62.5 & 0.975 & \cmark
        & 3.867 & 4.279 & \textbf{0.885} & \textbf{7.52}
        & 3.888 & 4.049 & 0.821 & \textbf{4.14} \\
      \rowcolor{bg_ours}
      \textbf{Ours} (37.5Hz)
        & 32k & 37.5 & 0.575 & \cmark
        & 3.914 & 4.346 & 0.849 & 9.57
        &3.924&4.161&0.774&4.81  \\
      \rowcolor{bg_ours}
      \textbf{Ours} (25.0Hz)
        & 32k & 25.0 & 0.375 & \cmark
        & 3.927 & 4.378 & 0.810 & 9.63
        &\textbf{3.937}&4.226&0.724&5.75 \\
      \rowcolor{bg_ours}
      \textbf{Ours} (18.75Hz)
        & 32k & 18.75 & 0.275 & \cmark
        & \textbf{3.942} & \textbf{4.426} & 0.777 & 10.18
        &3.932&\textbf{4.269}&0.684&6.50  \\
      
      \bottomrule
    \end{tabular}
  \vspace{-2mm}
\end{table*}

A key challenge in conditional flow-based audio generation is maintaining strict temporal correspondence between conditioning signals and latent trajectories. Inspired by F5-TTS~\cite{chen2025f5}, we enforce \emph{temporal alignment} by upsampling semantic tokens $\mathbf{z}_{\text{sem}}$, acoustic tokens $\mathbf{z}_{\text{ac}}$, and noisy VAE latents $\mathbf{u}_t$ to a shared frame rate and concatenating them frame-wisely:
\begin{equation}
\mathbf{c}_t =
\text{Concat}\!\left(
\text{Up}(\mathbf{u}_t),
\text{Up}(\mathbf{z}_{\text{sem}}),
\text{Up}(\mathbf{z}_{\text{ac}})
\right).
\end{equation}
This temporally aligned conditioning ensures that each generative step is informed by synchronized semantic structure and acoustic details, which is critical for stable flow matching and coherent long-form reconstruction.

Finally for the audio reconstruction, we follow rectified flow~\cite{liu2022flow} and define a linear probability path between the Gaussian noise $\mathbf{u}_0$ and the VAE latent $\mathbf{u}_1$:
\begin{equation}
\mathbf{u}_t = (1 - t)\mathbf{u}_0 + t \mathbf{u}_1, \quad t \in [0, 1],
\end{equation}
and train a DiT to predict the neural vector field $v_\theta$ using:
\begin{equation}
\mathcal{L}_{\text{FM}}
= \mathbb{E}_{t}
\left\|
v_\theta(\mathbf{c}_t, t)
- (\mathbf{u}_1 - \mathbf{u}_0)
\right\|_2^2.
\end{equation}

\paragraph{Training Objective of Stage 2.}
The acoustic encoder $E_{\text{ac}}(\cdot)$, its vector quantizer $Q_{\text{ac}}(\cdot)$, and the flow matching decoder $v_\theta(\cdot)$ are jointly optimized, while the semantic encoder trained from Stage 1 remains frozen. The overall objective is:
\begin{equation}
    \mathcal{L}_{\text{stage2}} =
    \mathcal{L}_{\text{FM}} + \lambda_{\text{VQ}}^{\text{ac}} \mathcal{L}_{\text{VQ}}^{\text{ac}},
\end{equation}
where $\lambda_{\text{VQ}}^{\text{ac}}$ balances the reconstruction fidelity and acoustic codebook stability.

\subsection{Inference}

At inference time, the input audio is first encoded into a set of semantic tokens and acoustic tokens. Conditioned on both representations, the generative decoder reconstructs clean latent representations via ODE-based sampling, implemented using the Euler method. The reconstructed latents are subsequently converted into time-domain waveforms by a VAE decoder~\cite{huang2023make}, followed by a BigVGAN-based neural vocoder~\cite {BIGVGAN}.

\section{Experiment}

We train our model on a large-scale compilation of approximately 180k hours of audio data from multiple sources, encompassing 100k hours of speech from VoxBox~\cite{voxbox}, 22k hours of environmental sounds from AudioSetCaps~\cite{bai2025audiosetcaps} and WavCaps~\cite{Wavcaps}, and 60k hours of instrumental music from VidMuse~\cite{tian2025vidmuse}, FMA~\cite{defferrard2016fma}, and in-house data collections.
We evaluate TeleSound-Codec on the \emph{speech}, \emph{sound}, and \emph{music} domains using a comprehensive set of objective and subjective metrics. These metrics assess perceptual quality, speaker similarity, and intelligibility for speech, as well as acoustic fidelity and aesthetic quality for sound and music. For reproducibility, detailed data statistics and implementation details are provided in our supplementary materials.

\subsection{Objective Evaluations}
\subsubsection{Evaluation Metrics}
\label{section:4.1}
For speech reconstruction, we conduct evaluation experiments on the \texttt{SeedTTS-test}~\cite{anastassiou2024seed} set. This dataset includes both Chinese and English speech. Thus, it enables assessment across different languages and phonetic characteristics.
We assess three complementary aspects: 1) Perceptual quality: mean opinion score (MOS) predicted by two objective speech quality models, DNSMOS~\cite{reddy2021dnsmos} and NISQA~\cite{mittag2021nisqa}, respectively; 2) Speaker similarity (SIM): cosine similarity between speaker embeddings extracted by a pretrained speaker verification model using the Wespeaker toolkit~\cite{wang2023wespeaker}. This metric reflects the preservation of speaker identity; 3) Intelligibility: we use character error rate (CER) for Chinese and word error rate (WER) for English. They are computed using a multilingual automatic speech recognition system, Whisper-medium~\cite{whisper}. 
These metrics jointly reflect whether the reconstructed speech is natural, intelligible, and speaker-consistent under extreme bitrate constraints.
For general sound reconstruction, we evaluate on the \texttt{AudioCaps-test}~\cite{kim2019audiocaps} set, which contains diverse real-world sound events. We focus on acoustic fidelity as is evaluated by the following metrics: 1) Fr\'echet audio distance (FAD): it measures the distance between the reconstructed and real audio distributions in a learned embedding space; 2) Kullback-Leibler divergence (KL): it evaluates distributional mismatch between the generated and ground-truth audio features; 3) Inception score (IS): it measures the clarity and diversity of acoustic events. All three metrics are computed based on audio features extracted by a pre-trained PaSST~\cite{koutini22passt} model that operates at a 32kHz sampling rate.
For music reconstruction, we conduct evaluations on the \texttt{MUSDB18-test}~\cite{rafii2017musdb18} dataset. 
We adopt metrics tailored to musical fidelity and aesthetics: 1) FAD and KL: they follow the same protocol as in the sound domain; 2) Aesthetic score (AES)~\cite{aes}: a composite metric of four perceptual dimensions, which are the content enjoyment (CE), content usefulness (CU), production complexity (PC), and production quality (PQ).

\begin{table*}[t]
  \caption{Objective evaluation in the sound and music domain. Evaluations are conducted on AudioCaps for the sound domain and MUSDB18 for the music domain.}
  \vspace{-2mm}
  \label{tab:combined_results_aes_grouped}
  \centering
  \footnotesize
    \begin{tabular}{lcccccccccccc}
      \toprule
      \multirow{3}{*}{\bf Method}
        & \multirow{3}{*}{\bf SR}
        & \multirow{3}{*}{\bf TR}
        & \multirow{3}{*}{\bf BR}
        & \multicolumn{3}{c}{\bf AudioCaps}
        & \multicolumn{6}{c}{\bf MUSDB18} \\
      \cmidrule(lr){5-7} \cmidrule(lr){8-13}
        & & & 
        & \multirow{2}{*}{\bf FAD $\downarrow$} & \multirow{2}{*}{\bf KL $\downarrow$} & \multirow{2}{*}{\bf IS $\uparrow$}
        & \multirow{2}{*}{\bf FAD $\downarrow$} & \multirow{2}{*}{\bf KL $\downarrow$} 
        & \multicolumn{4}{c}{\bf AES $\uparrow$} \\
      \cmidrule(lr){10-13}
        & & & 
        & & & 
        & & 
        & {\bf CE} & {\bf CU} & {\bf PC} & {\bf PQ} \\
      \midrule

      \rowcolor{bg_ref}
      \textcolor{gray}{Ground Truth}
        & \textcolor{gray}{-} & \textcolor{gray}{-} & \textcolor{gray}{-}
        & \textcolor{gray}{0.000} & \textcolor{gray}{0.000} & \textcolor{gray}{9.831} 
        & \textcolor{gray}{0.000} & \textcolor{gray}{0.000} & \textcolor{gray}{6.998} & \textcolor{gray}{7.685} & \textcolor{gray}{5.634} & \textcolor{gray}{7.629} \\
      
      \midrule

      Encodec~\cite{encodec}
        & 24k & 600 & 6.000 
        & 146.568 & 0.406 & 6.547 
        & 199.568 & 0.148 & 6.994 & 7.596 & 5.724 & 7.527 \\
      DAC~\cite{encodec}
        & 44k & 774 & 7.740 
        & 88.252 & 0.254 & 7.224 
        & 137.681 & 0.087 & 7.013 & 7.647 & 5.695 & 7.590 \\
      
      \midrule
      \multicolumn{13}{c}{\textit{Codecs operating at low bitrates}} \\
      \midrule

      \rowcolor{bg_bad}
      Encodec-RVQ1~\cite{encodec}
        & 24k & 75 & 0.750 
        & 646.595 & 2.922 & 2.393 
        & 603.160 & 0.594 & 5.344 & 6.413 & 5.221 & 6.467 \\
      \rowcolor{bg_bad}
      DAC-RVQ1~\cite{dac}
        & 44k & 86 & 0.860 
        & 476.125 & 2.546 & 3.151 
        & 373.077 & 0.298 & 6.406 & 6.968 & 5.828 & 6.847 \\

      \midrule

      \rowcolor{bg_sota}
      SemantiCodec~\cite{semanti}
        & 16k & 50 & 0.700 
        & 222.261 & 0.667 & 5.448 
        & 265.769 & 0.225 & 7.066 & 7.718 & 5.634 & 7.650 \\
      \rowcolor{bg_sota}
      WavTokenizer~\cite{ji2024wavtokenizer}
        & 24k & 40 & 0.480 
        & 323.436 & 1.712 & 3.691 
        & 302.325 & 0.267 & 7.071 & 7.682 & \textbf{5.800} & 7.657 \\
      \rowcolor{bg_sota}
      UniCodec~\cite{unicodec}
        & 24k & 75 & 1.050 
        & 374.545 & 1.297 & 3.802 
        & 424.029 & 0.332 & 6.784 & 7.439 & 5.544 & 7.378 \\
    \rowcolor{bg_sota}
      Moss-Audio-Tokenizer~\cite{moss}
        & 24k & 37.5 & 0.375 
        & 300.632 & 1.409 & 4.796 
        &372.570&0.401&6.750&7.538&5.595&7.535  \\
      
      \midrule
      
      \rowcolor{bg_ref}
      \textcolor{gray}{Ours VAE}
        & \textcolor{gray}{32k} & \textcolor{gray}{-} & \textcolor{gray}{-}
        & \textcolor{gray}{39.476} & \textcolor{gray}{0.136} & \textcolor{gray}{8.401} 
        & \textcolor{gray}{70.159} & \textcolor{gray}{0.048} & \textcolor{gray}{7.090} & \textcolor{gray}{7.798} & \textcolor{gray}{5.703} & \textcolor{gray}{7.782} \\

      \rowcolor{bg_ours}
      \textbf{Ours} (62.5Hz)
        & 32k & 62.5 & 0.975 
        & \textbf{96.926} & \textbf{0.463} & \textbf{7.069} 
        & \textbf{130.783} & \textbf{0.103} & \textbf{7.124} & 7.797 & 5.594 & 7.752 \\
        
      \rowcolor{bg_ours}
      \textbf{Ours} (37.5Hz)
        & 32k & 37.5 & 0.575 
        & 99.891 & 0.506 & 6.980 
        & 138.633 & 0.119 & 7.114 & \textbf{7.801} & 5.579 & 7.765 \\
        
      \rowcolor{bg_ours}
      \textbf{Ours} (25.0Hz)
        & 32k & 25.0 & 0.375 
        & 106.378 & 0.558 & 6.910 
        & 148.183 & 0.126 & 7.106 & 7.798 & 5.561 & 7.774 \\
        
      \rowcolor{bg_ours}
      \textbf{Ours} (18.75Hz)
        & 32k & 18.75 & 0.275 
        & 115.663 & 0.626 & 6.719 
        & 158.489 & 0.146 & 7.090 & 7.798 & 5.545 & \textbf{7.782} \\
      
      \bottomrule
    \end{tabular}
  \vspace{-2mm}
\end{table*}

\subsubsection{Results and Analysis}

We evaluate speech reconstruction performance on the \texttt{SeedTTS-test} set. As detailed in Table~\ref{tab:seedtts_results}, TeleSound-Codec reaches the state-of-the-art performance for ultra-low bitrate coding.
Operating at a bitrate as low as 0.275kbps, our model achieves high perceptual quality and intelligibility across both Chinese and English subsets, with only a slight loss in speaker similarity. This performance represents a significant advancement over early multi-codebook baselines such as \emph{Encodec} and \emph{DAC}. When reducing the codebook size for lower bitrates, the compared methods tend to suffer from severe quality degradation and even perceptual collapse. Moreover, we also outperform recent strong low-bitrate contenders like \emph{SemantiCodec}, \emph{WavTokenizer}, \emph{UniCodec}, and \emph{Moss-Audio-Tokenizer} in both perceptual quality and intelligibility, even when these baselines are allowed higher bandwidths. Our language-grounded design ensures robust content accuracy, yielding lower WER/CER than the compared methods that lack explicit linguistic alignment.

Interestingly, we observe a slight counter‑intuitive pattern in Table~\ref{tab:seedtts_results} where lower bitrates occasionally yield marginally higher perceptual scores. We posit that as the acoustic token rate decreases, the model is forced to discard fine-grained spectral details, which often contain background noise or recording artifacts. Consequently, the generative decoder relies more heavily on the clean, language-grounded semantic priors to synthesize speech. This process effectively functions as a denoising mechanism, resulting in \emph{cleaner} and more intelligible speech reconstruction over faithful but noisy high-fidelity reproduction.

For the sound domain, Table~\ref{tab:combined_results_aes_grouped} reports the objective evaluation on the \texttt{AudioCaps-test} dataset. In this domain, which often involves complex environmental textures, our method demonstrates exceptional robustness compared to recent strong baselines.
Acoustic codecs without semantic supervision, such as \emph{Encodec} and \emph{DAC}, typically rely on relatively high token rates to preserve environmental realism. If aggressively compressed, their perceptual quality degrades substantially.
TeleSound-Codec breaks this limitation by maintaining realistic acoustic rendering at ultra-low bitrates. Compared with both semantic codecs and purely acoustic baselines that operate at higher bitrates, our method consistently achieves much lower distortion metrics and a higher inception score, indicating superior preservation of acoustic event realism. 
This behavior confirms that the proposed semantic-first tokenization captures high-level sound event concepts that can be faithfully rendered by the strong generative decoder, even under an extremely tight information bottleneck.

For the music reconstruction on the \texttt{MUSDB18-test} set, TeleSound-Codec demonstrates clear advantages in both acoustic fidelity and aesthetic preservation, as summarized in Table~\ref{tab:combined_results_aes_grouped}. Across a wide range of bitrates, our method achieves lower signal distortion while maintaining higher aesthetic quality. Even at 0.275kbps, it preserves low FAD and KL divergence and delivers competitive or superior AES scores compared to baseline methods operating at higher bitrates. These results indicate that TeleSound-Codec remains effective for music coding at high compression ratios.

\subsubsection{Flexible Bitrate Control}
Our framework enables flexible bitrate control via the adaptive pooling stride of the acoustic encoder. As shown in Table~\ref{tab:seedtts_results} and Table~\ref{tab:combined_results_aes_grouped}, we evaluate four distinct bitrates ranging from 0.975kbps down to 0.275kbps. The results demonstrate a smooth quality-bitrate trade-off.
This robustness is underpinned by our language-grounded semantic tokens, which establish a solid content foundation invariant to bitrate changes. Even at the lowest bandwidths, these tokens ensure semantic intelligibility and structural coherence. Building upon this stable ground, the rate-adaptive acoustic tokens efficiently capture the necessary residual details, allowing the powerful flow matching decoder to synthesize high-fidelity waveforms. Consequently, reducing the bitrate from 0.975kbps to 0.375kbps results in negligible perceptual loss, confirming the efficacy of this layered representation where semantic stability supports flexible acoustic modeling. Consequently, this layered representation ensures that the model maintains essential usability at extremely low bitrates, while scaling to superior audio fidelity as bandwidth increases.

\subsection{Subjective Evaluations}
\subsubsection{Experimental Setup}
We conduct subjective listening tests following the MUSHRA~\citep{MUSHRA} protocol\footnote{\url{https://github.com/audiolabs/webMUSHRA}} to evaluate the perceptual quality of different audio tokenization methods across the \emph{speech}, \emph{sound}, and \emph{music} domains. This evaluation is designed to assess the overall perceptual quality of reconstructed audio, including fidelity, naturalness, and the presence of audible artifacts under different tokenization strategies.
For the speech domain, we randomly select $5$ utterances from the \texttt{SeedTTS-test} set for each of the Chinese and English subsets, resulting in a total of $10$ speech samples. For the sound domain, we randomly select $10$ samples from the \texttt{AudioCaps-test} set. For the music domain, we randomly select $10$ music excerpts from the \texttt{MUSDB18-test} set.
All audio samples are loudness-normalized before evaluation to ensure a fair comparison across different systems. In each MUSHRA trial, listeners are presented with a reference signal, a hidden reference, and multiple test systems corresponding to different audio tokenization methods. The order of the test systems is randomized independently for each trial to avoid any bias introduced by a fixed presentation order. Listeners are asked to rate each stimulus on a continuous scale from 0 to 100 according to its overall perceptual quality, where the reference represents the upper bound of perceptual quality. The evaluation process follows the standard protocol defined by the ITU~\cite{MUSHRA}.

\subsubsection{Results and Analysis}

Table~\ref{tab:mushra} summarizes the MUSHRA results (mean $\pm$ 95\% CI) for different datasets and methods.

\begin{table}[!t]
\centering
\footnotesize
\caption{MUSHRA results (mean $\pm$ 95\% confidence interval) across different audio domains.}
\vspace{-2mm}
\label{tab:mushra}
\begin{tabular}{lccc}
\toprule
\textbf{Method} & \textbf{Speech} & \textbf{Sound} & \textbf{Music} \\
\midrule
Reference 
& 94.44 $\pm$ 1.51 
& 94.46 $\pm$ 1.33 
& 94.31 $\pm$ 1.37 \\
\midrule

UniCodec ~\citep{unicodec}
& 79.59 $\pm$ 3.41 
& 47.83 $\pm$ 4.33 
& 56.74 $\pm$ 4.26 \\

SemantiCodec~\citep{semanti} 
& 60.19 $\pm$ 4.02 
& 60.09 $\pm$ 4.28 
& 63.26 $\pm$ 3.92 \\

WavTokenizer~\citep{ji2024wavtokenizer}
& 64.43 $\pm$ 4.45 
& 45.71 $\pm$ 4.31 
& 49.14 $\pm$ 4.39 \\
\textbf{Ours} 
& \textbf{84.86} $\pm$ 2.58 
& \textbf{78.23} $\pm$ 3.05 
& \textbf{81.21} $\pm$ 3.01 \\

\bottomrule
\end{tabular}
\vspace{-4mm}
\end{table}


As shown in Table~\ref{tab:mushra}, the proposed method consistently achieves the highest MUSHRA scores among all evaluated methods across speech, sound, and music domains, while remaining close to the reference signal. The subjective evaluation results are consistent with the trends observed in the objective metrics, which further supports the validity of the objective evaluation results.
\subsection{Integration with Language-Model Based Generation}
\label{sec:downstream}

To validate TeleSound-Codec in downstream tasks, we build a zero-shot text-to-speech (TTS) system using Qwen3-0.6B \cite{yang2025qwen3}. Accommodating our dual-stream design, we flatten the 12.5Hz semantic and 50Hz acoustic tokens into an interleaved sequence: $\mathbf{T}_{\text{speech}} = [s_1, a_{1}, a_{2}, a_{3}, a_{4}, s_2, \dots]$. The model is fine-tuned using autoregressive cross-entropy loss and compared against low-bitrate baselines, WavTokenizer \cite{ji2024wavtokenizer}, and UniCodec \cite{unicodec}. To demonstrate the learning efficiency advantage of the proposed method in downstream tasks, we uniformly limit the training steps of all methods to 200k.

\begin{table}[!t]
  \caption{Zero-shot in-context learning based text-to-speech evaluation on SeedTTS-ZH and SeedTTS-EN.}
  \vspace{-2mm}
  \label{tab:seedtts_results_slim}
  \centering
  \footnotesize
  \resizebox{\columnwidth}{!}{
    \setlength{\tabcolsep}{1pt}
    \begin{tabular}{lcccccccc}
      \toprule
      \multirow{3}{*}{\bf Codec}
        & \multicolumn{4}{c}{\bf SeedTTS-ZH}
        & \multicolumn{4}{c}{\bf SeedTTS-EN} \\
      \cmidrule(lr){2-5} \cmidrule(lr){6-9}
        & \multicolumn{2}{c}{\bf MOS$\uparrow$} & \multirow{2}{*}{\bf SIM$\uparrow$} & \multirow{2}{*}{\bf CER (\%)$\downarrow$}
        & \multicolumn{2}{c}{\bf MOS$\uparrow$} & \multirow{2}{*}{\bf SIM$\uparrow$} & \multirow{2}{*}{\bf WER (\%)$\downarrow$} \\
      \cmidrule(lr){2-3} \cmidrule(lr){6-7}
        & {\bf DNS} & {\bf NISQA} & & 
        & {\bf DNS} & {\bf NISQA} & & \\
      \midrule
      WavTokenizer~\cite{ji2024wavtokenizer}
        &3.727&3.699&0.708&30.32 
        &3.738&3.731&\textbf{0.666}&21.70  \\
      UniCodec~\cite{unicodec}
        &\textbf{3.753}&\textbf{4.097}&0.711&25.74 
        &\textbf{3.838}&\textbf{4.153}&0.663&49.32  \\
      \textbf{Ours}
        &3.669&4.051&\textbf{0.712}&\textbf{8.75} 
        &3.636&3.774&0.612&\textbf{3.34}  \\
      \bottomrule
    \end{tabular}
    }
    \vspace{-2mm}
\end{table}

As shown in Table~\ref{tab:seedtts_results_slim}, our method achieves significantly lower CER (8.75\%) and WER (3.34\%) than the baselines, alongside competitive perceptual quality. This highlights that our semantic-first design significantly simplifies the LLM's text-to-audio mapping compared to acoustic-centered approaches.

To further assess modeling difficulty of audio tokens from different codecs, we compute the normalized perplexity (PPL) \cite{wang2025audiocodecbench}, defined as $\text{PPL} = \exp(\mathcal{L}_{CE}) \times \frac{1024}{S_{cb}}$, where $\mathcal{L}_{CE}$ is the cross-entropy loss and $S_{cb}$ is the codebook size.

\begin{table}[!t]
\centering
\footnotesize
\caption{Comparison of perplexity (PPL) on SeedTTS-ZH and SeedTTS-EN Datasets}
\vspace{-2mm}
\label{tab:ppl_comparison}
\begin{tabular}{lcc}
\toprule
\textbf{Method} & \textbf{SeedTTS-ZH} & \textbf{SeedTTS-EN} \\
\midrule
WavTokenizer    & 55.036 & 65.878 \\
UniCodec        & 19.727 & 21.043 \\
\midrule
Ours (semantic) & 0.808  & 1.158  \\
Ours (acoustic) & 5.886  & 6.959  \\
\bottomrule
\end{tabular}
\vspace{-4mm}
\end{table}

Table~\ref{tab:ppl_comparison} demonstrates that our disentangled approach drastically lowers overall PPL. The semantic tokens exhibit extremely low PPL due to their highly deterministic mapping from text, while acoustic tokens efficiently capture the remaining stochastic variations (e.g., prosody, timbre). This divide-and-conquer strategy substantially reduces the modeling complexity for the LLM.
\begin{figure*}[t]
    \centering
    \includegraphics[width=0.8\linewidth]{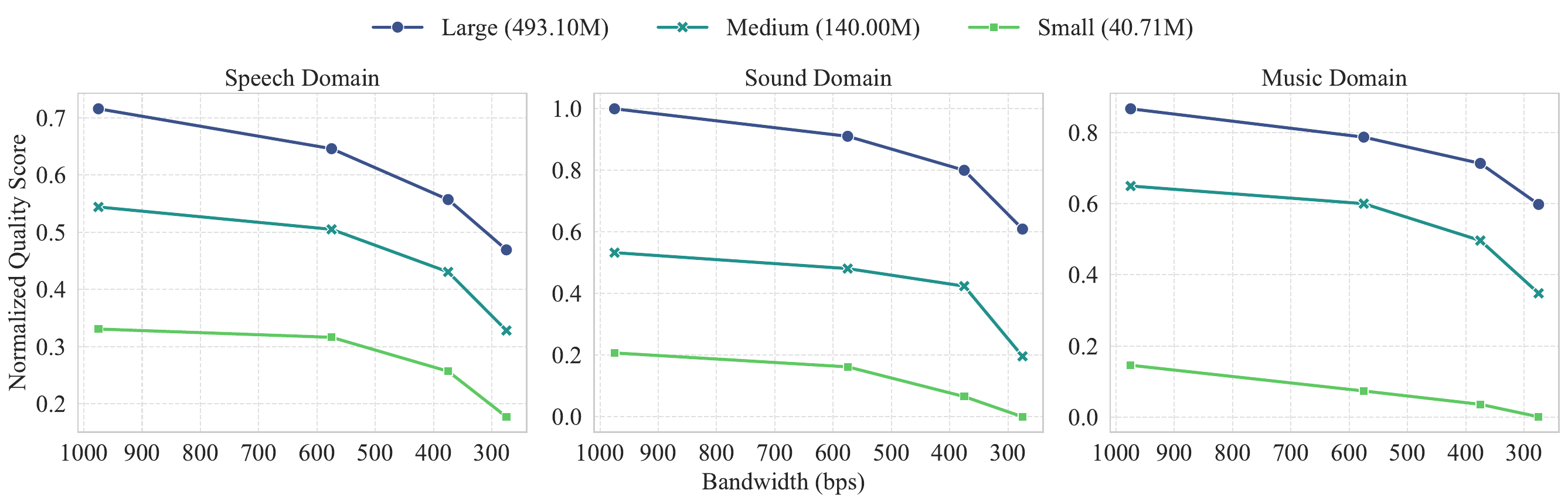}
    \caption{Illustration of the model performance varying with the bandwidth of different-sized models in the speech, sound, and music domains.}
    \Description{Illustration of the model performance varying with the bandwidth of different-sized models in the speech, sound, and music domains.}
    \label{fig:tradeoff}
\end{figure*}

\begin{table*}[t]
\centering
\footnotesize
\caption{Ablation study. For better visualization, the MOS score is calculated as a mean of DNS and NISQA; the AES score is calculated as a mean of all four AES metrics (CE/CU/PC/PQ).}
\vspace{-2mm}
\label{tab:ablation_colored}
\setlength{\tabcolsep}{3.5pt}

\begin{tabular}{lcccccccccccccc}
\toprule
\multirow{3}{*}{\bf Config.} 
& \multirow{3}{*}{\bf TR} 
& \multirow{3}{*}{\bf BR} 
& \multicolumn{6}{c}{\bf SeedTTS} 
& \multicolumn{3}{c}{\bf AudioCaps} 
& \multicolumn{3}{c}{\bf MUSDB18} \\
\cmidrule(lr){4-9} \cmidrule(lr){10-12} \cmidrule(lr){13-15}
& & 
& \multicolumn{3}{c}{\bf ZH} 
& \multicolumn{3}{c}{\bf EN} 
& \multirow{2}{*}{\bf FAD $\downarrow$} & \multirow{2}{*}{\bf KL $\downarrow$} & \multirow{2}{*}{\bf IS $\uparrow$} 
& \multirow{2}{*}{\bf FAD $\downarrow$} & \multirow{2}{*}{\bf KL $\downarrow$} & \multirow{2}{*}{\bf AES $\uparrow$} \\
\cmidrule(lr){4-6} \cmidrule(lr){7-9}
& & 
& {\bf MOS $\uparrow$} & {\bf SIM $\uparrow$} & {\bf CER (\%) $\downarrow$} 
& {\bf MOS $\uparrow$} & {\bf SIM $\uparrow$} & {\bf WER (\%) $\downarrow$} 
& & & & & & \\
\midrule

\multirow{4}{*}{\bf Full config.}
& 62.5 & 0.975 &  4.073 &  \textbf{0.885} &  \textbf{7.52} &  3.969 &  \textbf{0.821} &  \textbf{4.14} &  \textbf{96.926} &  \textbf{0.463} &  \textbf{7.069} &  \textbf{130.783} &  \textbf{0.103} &  \textbf{7.067} \\
& 37.5 & 0.575 &  4.130 &  0.849 &  9.57 &  4.043 &  0.774 &  4.81 &  99.891 &  0.506 &  6.980 &  138.633 &  0.119 &  7.065 \\
& 25.0 & 0.375 &  4.153 &  0.810 &  9.63 &  4.082 &  0.724 &  5.75 &  106.378 &  0.558 &  6.910 &  148.183 &  0.126 &  7.060 \\
& 18.75 & 0.275 &  4.184 &  0.777 &  10.18 &  4.101 &  0.684 &  6.50 &  115.663 &  0.626 &  6.719 &  158.489 &  0.146 &  7.054 \\
\midrule

\multirow{4}{*}{Acoustic only}
& 50 & 0.800 & 4.108 & 0.838 & 45.54 & 3.998 & 0.757 & 42.96 & 141.973 & 0.870 & 5.776 & 150.513 & 0.129 & 7.062 \\
& 25 & 0.400 & 4.113 & 0.756 & 130.48 & 4.040 & 0.678 & 84.59 & 164.013 & 1.146 & 5.415 & 167.526 & 0.171 & 7.051 \\
& 12.5 & 0.200 & 3.998 & 0.669 & 334.13 & 3.964 & 0.607 & 122.99 & 200.609 & 1.692 & 4.679 & 198.323 & 0.251 & 7.040 \\
& 6.25 & 0.100 & 3.703 & 0.601 & 712.57 & 3.716 & 0.566 & 179.49 & 228.561 & 2.264 & 4.247 & 238.304 & 0.335 & 6.949 \\
\midrule

Semantic only
& 12.5 & 0.175 & \textbf{4.221} & 0.730 & 10.81 & \textbf{4.126} & 0.636 & 7.22 & 134.402 & 0.803 & 6.620 & 185.532 & 0.204 & 7.014 \\
\quad\emph{-- w/o LM}
& 12.5 & 0.175 &4.015&0.705&99.19&3.804&0.598&95.24&153.759&1.137&5.904&192.698&0.243&6.970 \\
\bottomrule
\end{tabular}
\vspace{-2mm}
\end{table*}

\subsection{Trading Computation for Bandwidth}
We investigate the scaling properties of TeleSound-Codec by training decoders of varying sizes: Small (40.71M), Medium (140.00M), and Large (493.10M). To ensure a fair comparison, we use an identical encoder for all models and freeze it during training. We only optimize the decoder parameters. Each decoder is trained for 300,000 steps until the training loss plateaus. Figure~\ref{fig:tradeoff} illustrates the normalized quality scores across speech, sound, and music domains as a function of the bandwidth\footnote{For visualization, we normalized all metrics to a [0, 1] scale and calculated the mean score for each domain to illustrate the trade-off. Please refer to our supplementary materials for the complete raw experimental data and the detailed normalization methodology.}. We observe a consistent \emph{scaling trend} behavior: for a fixed bandwidth, increasing the model size yields higher reconstruction quality. Crucially, the Large model at lower bitrates (e.g., 0.3kbps) often matches or exceeds the performance of the Small model at higher bitrates (e.g., 0.6kbps). This confirms our core hypothesis that the power of generative computing can compensate for the degradation of the transmission bandwidth, allowing high-fidelity decoding at ultra-low bitrates.

\subsection{Ablations}
\label{sec:ablations}
To validate the contributions of the semantic and acoustic components, we conduct an ablation study presented in Table~\ref{tab:ablation_colored}.
Removing the semantic stream (``Acoustic only") leads to a catastrophic degradation in speech WERs and CERs, as well as a great loss in audio ISs. This confirms that semantic tokens are essential for preserving content intelligibility; 
Removing the acoustic stream (``Semantic only") causes an acoustic distortion, as reflected by the decreased speaker similarity in the speech domain, as well as increased FAD and KL in the sound and music domain. This indicates that acoustic tokens are necessary for capturing fine-grained residual acoustic details, which are essential for high-fidelity audio reconstruction;  Removing LM supervision (``w/o LM'') causes substantial degradation in both probing accuracy and reconstruction quality across sound, music, and speech, demonstrating the critical role of language-model-based semantic grounding in shaping informative latent representations and facilitating downstream generation.

The ablation results also indicate that, at extremely low bitrates, language-grounded semantic tokens can even outperform more than 4-times higher-bitrate acoustic-only tokens in terms of reconstruction quality, demonstrating that the strong grounding provided by linguistically aligned representations significantly facilitates audio reconstruction. Based on this semantic foundation, the decoder further reconstructs fine-grained acoustic residuals captured by the dedicated acoustic encoder, enabling high-fidelity synthesis while preserving content consistency. This validates the effectiveness of the proposed semantic-acoustic dual-stream design for achieving both intelligibility and perceptual quality under stringent bitrate constraints.

\section{Conclusion}
In this work, we introduced TeleSound-Codec, a semantic-acoustic dual-stream neural audio codec designed for high-fidelity general audio coding in extremely low bitrate regimes. 
By explicitly grounding discrete semantic representations with large language model supervision, our approach maintains semantic integrity under aggressive compression. Building upon these language-grounded semantics, we further propose a scalable flow-matching-based generative decoder. The decoder is jointly conditioned on complementary dual-stream tokens, including semantic tokens that provide high-level content guidance and acoustic tokens that capture fine-grained acoustic details. This design preserves both semantic integrity and acoustic fidelity across speech, music, and general sounds at ultra-low bitrates.
Extensive experiments demonstrate that TeleSound-Codec consistently outperforms both acoustic and semantic baselines at bitrates below 1kbps, achieving strong perceptual quality and intelligibility even at 0.275kbps. Moreover, our results reveal a clear and practical compute-bandwidth trade-off: increasing the capacity of the generative model significantly improves the reconstruction quality at a fixed bandwidth. Despite these advantages, we acknowledge that the computational cost and non-causal nature of the proposed model limit its applicability in real-time streaming scenarios. Based on the observed trade-off between computation and performance, an important direction for future work is to develop more efficient and streaming-friendly variants that balance latency and quality, enabling deployment in real-time communication and interactive audio applications.
\balance
\bibliographystyle{ACM-Reference-Format}
\bibliography{ref}

\clearpage
\section{Implementation Details}
\label{append:detail}
\begin{table}[htb]
\centering
\footnotesize
\caption{Statistics of the training datasets.}
\label{tab:dataset_statistics}
\resizebox{\columnwidth}{!}{%
\begin{tabular}{l l r r}
\toprule
\textbf{Domain} & \textbf{Source} & \textbf{\# Samples} & \textbf{Duration (h)} \\
\midrule
Speech 
& VoxBox 
& 48,964,663 
& 104,276 \\
\midrule
\multirow{2}{*}{Sound} 
& AudioSetCaps 
&  5,740,939
&  15,947\\
& Internal Data 
&  2,919,148
&  6,497\\
\midrule
\multirow{4}{*}{Music} 
& VidMuse 
& 337,236
& 15,732 \\
& FMA 
& 13,577 
&  714 \\
& Internal Data 
&  730,816
&  42,684\\
\bottomrule
\end{tabular}
}
\end{table}
\paragraph{Training Datasets.}
We train our model on a large-scale compilation of approximately 180k hours of audio data, encompassing speech, environmental sounds, and instrumental music.
The speech data is sourced from the VoxBox dataset~\cite{voxbox}, which provides multi-dimensional attribute labels, including language, age, gender, emotion, pitch, and speaking rate. Based on these attributes, we design a predefined template function that combines the structured labels and converts them into natural language descriptions, which serve as speaker-level captions aligned with the corresponding speech signals.
The sound data consist of two parts: 1) the full AudioSetCaps dataset~\cite{bai2025audiosetcaps}, and 2) approximately 3 million high-quality sound effect samples collected internally. The internal sound data are annotated with fine-grained semantic descriptions generated by Gemini-2.5 Pro~\cite{comanici2025gemini}.
The music data are collected from a combination of public and internal sources. Specifically, VidMuse~\cite{tian2025vidmuse} and Free Music Archive (FMA)~\cite{defferrard2016fma} are open-source datasets. The open-source music datasets are annotated using Kimi-Audio~\cite{kimiaudio}, whereas the internal music data are annotated with Gemini-2.5 Pro.
Detailed statistics and scale information for each category of training data are reported in Table~\ref{tab:dataset_statistics}. To mitigate data imbalance across domains during training, we apply 3 times and 6 times resampling to the sound and music data, respectively, which helps enhance the model’s ability to represent non-speech audio.

\paragraph{Detailed Model Configurations.}
Our model is trained in two independent stages. In \textbf{Stage 1}, we train a semantic encoder with explicit textual supervision to obtain discrete, linguistically grounded semantic tokens. The textual supervision is provided by a Qwen3-1.7B language model head~\cite{yang2025qwen3}, which is fine-tuned using LoRA~\cite{hu2022lora} with rank 16, following the task configurations summarized in Table~\ref{tab:audio_tasks}. The semantic encoder is initialized from the Whisper-large V3~\cite{whisper} audio encoder and is fully trainable during this stage. Token representations from the 24th encoder layer are downsampled by a factor of 4 along the temporal dimension and discretized using vector quantization with a codebook size of 16,384. This results in a discrete semantic token sequence with a frame rate of 12.5Hz and an effective bitrate of 14bits per token, representing a bitrate of 175bps. To encourage temporal consistency of the learned representations, we additionally leverage ATST-Frame~\cite{li2024self} to extract self-supervised features for temporal regularization. The loss scalers controlling the contributions of the quantization loss and the temporal regularization loss are set to $\lambda_{\text{VQ}}^{\text{sem}} = 10.0$ and $\lambda_{\text{temp}} = 1.0$, respectively. During training, we adopt a dynamic batching strategy that samples approximately 1,600 seconds of audio per optimization step across 8 H100 GPUs. The learning rate is linearly warmed up from 0 to $1\times10^{-4}$ over the first 20k steps and then annealed to zero using a cosine schedule. The semantic encoder is trained for a total of 500k steps.

In \textbf{Stage 2}, the semantic encoder is frozen, and we jointly train an acoustic encoder and a diffusion Transformer-based decoder for high-fidelity acoustic reconstruction. The acoustic encoder is implemented as a 12-layer Transformer with a hidden dimension of 1024, a feed-forward dimension of 3072, and 16 attention heads. It takes 128-dimensional log-Mel spectrograms extracted from 32~kHz audio at a frame rate of 50~Hz as input. To support variable-rate acoustic tokenization, the encoder applies dynamic temporal downsampling with factors of 1, 2, 4, and 8, followed by vector quantization with a codebook size of 65,536, producing discrete acoustic token sequences with token rates of 50, 25, 12.5, and 6.25~Hz, where each token is represented using 16 bits. The acoustic decoder is based on a diffusion Transformer architecture trained with a rectified flow matching objective. The DiT backbone consists of 24 Transformer blocks with a hidden dimension of 1024 and a feed-forward dimension of 3072, employing rotary positional embeddings (RoPE)~\cite{rope} for long-range temporal modeling. Each block adopts RMSNorm~\cite{zhang2019root} and a LLaMA-style feed-forward network~\cite{touvron2023llama} with SwiGLU activation~\cite{shazeer2020glu}, and injects the diffusion time step $t$ via adaptive layer normalization (AdaLN)~\cite{peebles2023scalable}. The decoder operates on a VAE~\cite{huang2023make} latent space of dimension 64 and is jointly conditioned on the semantic tokens and the acoustic encoder outputs. The loss scaler for the acoustic quantization loss is set to $\lambda_{\text{VQ}}^{\text{ac}} = 10.0$. Classifier-free guidance~\cite{cfg} is applied during training with a drop rate of 0.1 by randomly zeroing out both semantic and acoustic tokens. We adopt a dynamic batching strategy that samples approximately 2,560 seconds of audio per optimization step across 8 H100 GPUs. We first jointly train the acoustic encoder and the DiT decoder with an initial learning rate of $1\times10^{-4}$, which is annealed to zero using a cosine schedule over 300k steps. The acoustic encoder is then frozen, and the DiT decoder is further fine-tuned for an additional 200k steps using a fixed learning rate of $1\times10^{-5}$. All training is conducted using BF16 mixed-precision arithmetic and the AdamW~\cite{adamw} optimizer with betas set to (0.9, 0.95) and a weight decay of 0.01.

\section{Implementation Details of LLM-based Generation}
\label{append:downstream}

\begin{figure*}[ht]
    \centering
    \includegraphics[width=0.8\linewidth]{tts_illustrator_v1.pdf}
    \caption{Overview of a text-to-speech framework integrating our TeleSound-Codec with a language model. (a) illustrates the speech tokenizer, where the input speech is disentangled into semantic and acoustic tokens and flattened into an interleaved sequence. (b) depicts the text-speech language model for conditional generation. Dashed lines indicate the autoregressive prediction of speech tokens. (c) illustrates the speech decoder conditioning on the semantic and acoustic tokens, which utilizes a diffusion transformer and a vocoder to reconstruct the final speech waveform.}
    \Description{Illustration of language-model based generation pipeline}
    \label{fig:tts_illustrator}
\end{figure*}
\begin{figure*}[t]
    \centering
    \begin{subfigure}[t]{0.33\linewidth}
        \centering
        \includegraphics[width=\linewidth]{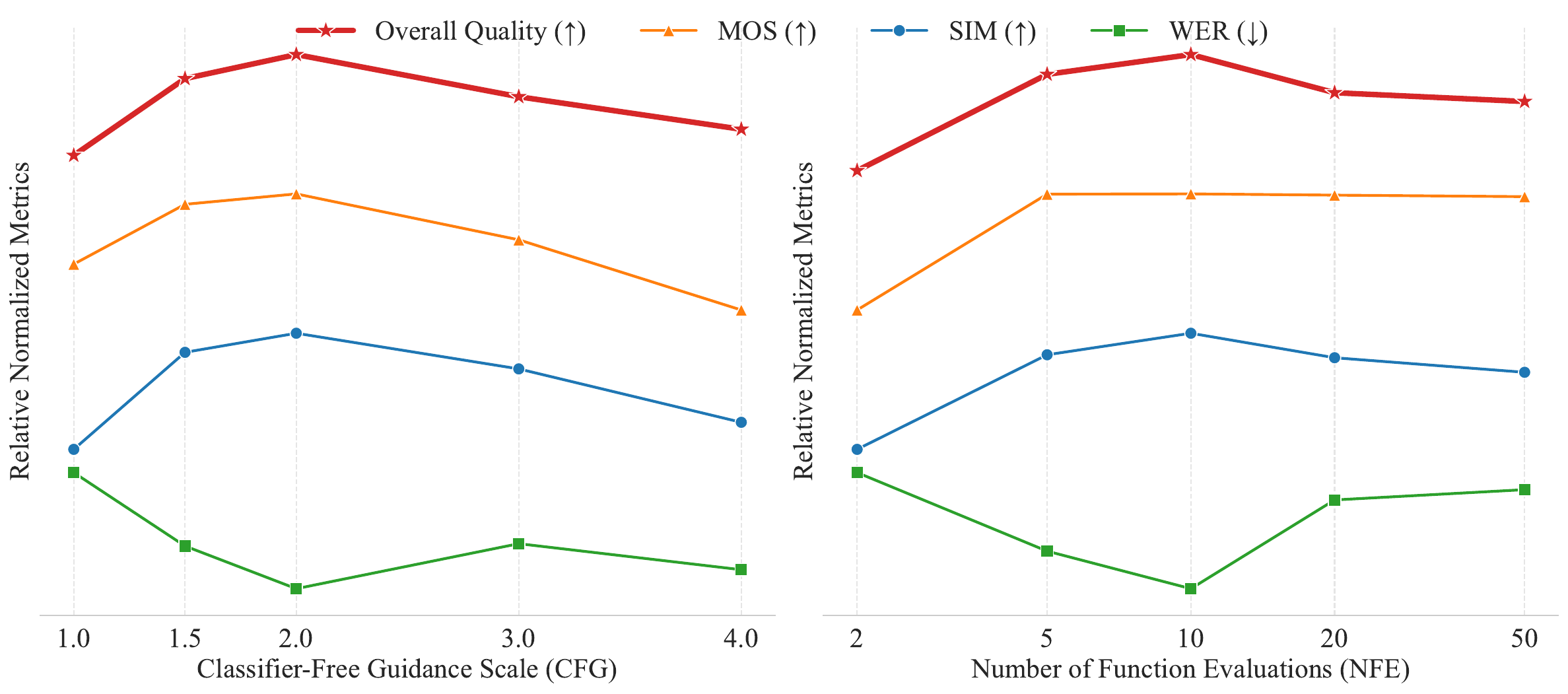}
        \caption{Speech domain (LibriTTS-eval-clean)}
    \end{subfigure}
    \begin{subfigure}[t]{0.33\linewidth}
        \centering
        \includegraphics[width=\linewidth]{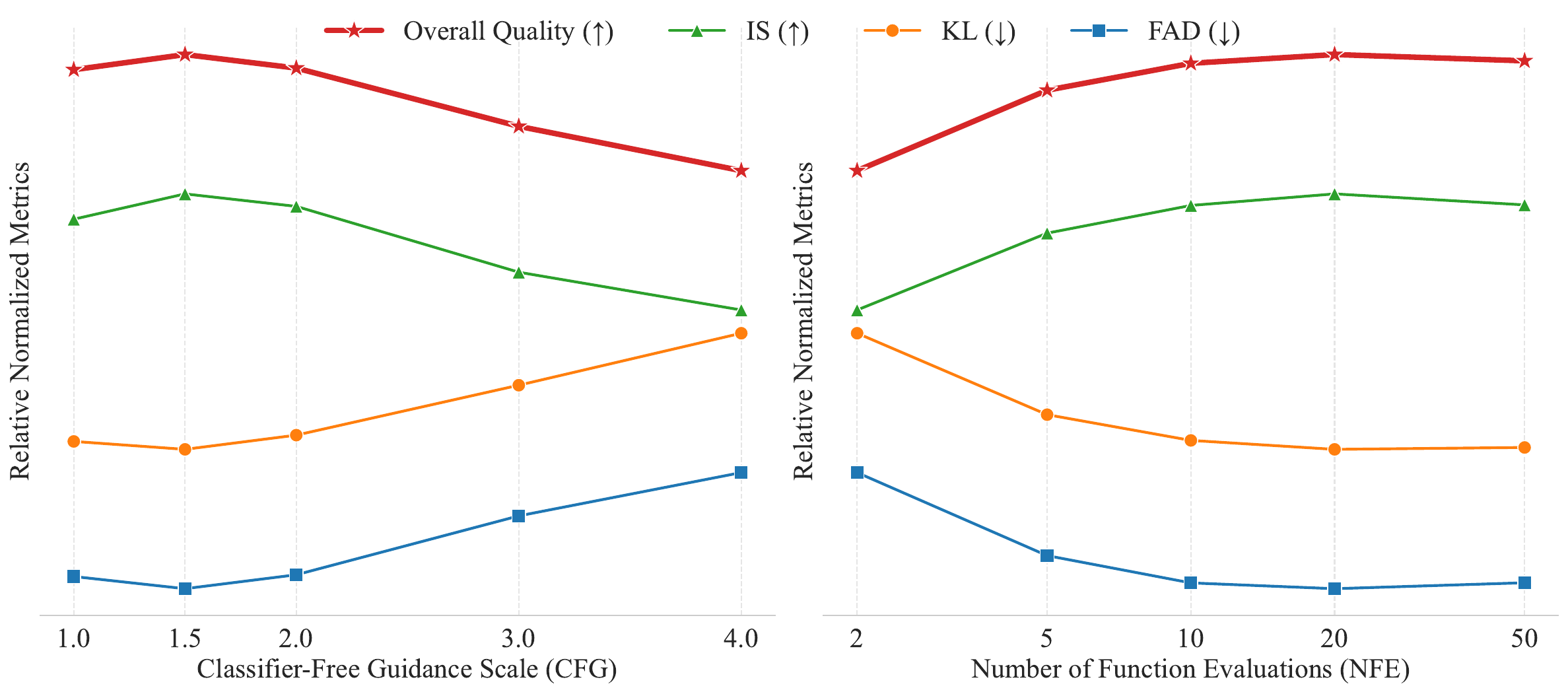}
        \caption{Sound domain (AudioCaps-val)}
    \end{subfigure}
    \begin{subfigure}[t]{0.33\linewidth}
        \centering
        \includegraphics[width=\linewidth]{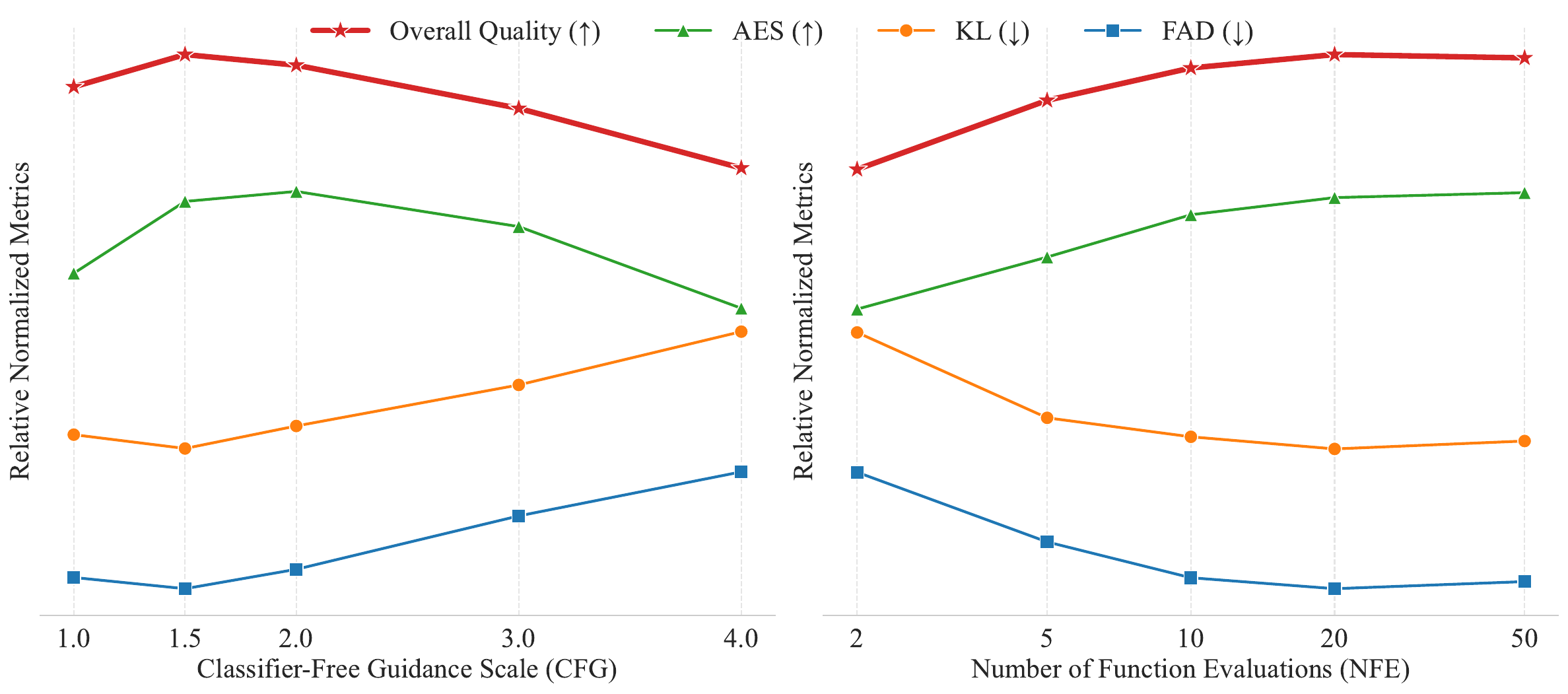}
        \caption{Music domain (MUSDB18-val)}
    \end{subfigure}
    
    \caption{Hyperparameter ablation across three audio domains. We analyze the effect of CFG scale and NFEs on generation quality.}
    \Description{Hyperparameter ablation across three audio domains.}
    \label{fig:ablation_all}
\end{figure*}

We present the process of integrating discrete audio tokens into an LLM-based audio generation pipeline in Figure~\ref{fig:tts_illustrator}. Specifically, we adopt Qwen3-0.6B \cite{yang2025qwen3} as our LLM backbone model and perform full fine-tuning initialized with pre-trained weights. To accommodate the dual-stream nature of TeleSound-Codec, we perform a semantic-acoustic interleaved sequence modeling strategy. Specifically, we employ the 1.8B configuration of TeleSound-Codec operating in the 62.5Hz inference mode. Given that the semantic token frame rate is 12.5Hz and the acoustic token frame rate is 50Hz, there exists a 1:4 temporal correspondence. Consequently, we flatten the discrete streams into a repeating pattern of \emph{one semantic token followed by four acoustic tokens}:
\begin{equation}
    \mathbf{T}_{\text{speech}} = [s_1, a_{1}, a_{2}, a_{3}, a_{4}, s_2, a_{5}, \dots].
\end{equation}
During the decoding stage, the generated sequence is separated back into semantic and acoustic streams, which are then fed into the TeleSound-Codec decoder to reconstruct the waveform. For baselines, we select two representative unified audio codecs: UniCodec~\cite{unicodec} and WavTokenizer~\cite{ji2024wavtokenizer}. Since these methods operate as single-codebook codecs, their corresponding speech token sequences are consequently represented in a simple flattened format.

We utilize only the speech subset listed in Table~\ref{tab:dataset_statistics} for training. The model is optimized using the standard autoregressive cross-entropy loss, computed exclusively on the speech tokens. We train the model on 2 GPUs using a dynamic batching strategy, capping the maximum number of tokens per device at 9,000. Additionally, the gradient accumulation count is set to 2. We use the AdamW optimizer with an initial learning rate of $1\text{e}^{-4}$. All experiments are performed using BF16 mixed precision for a maximum of 200k steps. The learning rate schedule includes a warm-up phase of 2,500 steps, followed by a constant phase of 75,000 steps, and finally a power function decay.

\section{Hyperparameter Selections for Flow Matching}
\label{sec:appendix_ablation}

We analyze the impact of key hyperparameters in flow matching, including the classifier-free guidance (CFG) scale and the number of function evaluations (NFEs), across three audio domains: speech, sound, and music.
We illustrate all the results in Figure~\ref{fig:ablation_all}

In the \emph{speech} domain, we evaluate on 200 randomly sampled clips from \texttt{LibriTTS-eval-clean}. As shown in Figure~\ref{fig:ablation_all} (a), increasing the CFG scale initially improves quality, while values beyond $\lambda = 2.0$ lead to clear degradation. The performance saturates at around 10 NFEs; In the \emph{sound} domain, experiments are conducted on the \texttt{AudioCaps-val} dataset. As shown in Figure~\ref{fig:ablation_all} (b), the aggregated metrics (IS, KL, and FAD) reach a stable optimum at $\lambda = 1.5$ with $\text{NFE} = 20$; In the \emph{music} domain, we evaluate on a held-out subset of 95 samples from \texttt{MUSDB18}. As shown in Figure~\ref{fig:ablation_all} (c), the best performance is achieved at $\lambda = 1.5$ and $\text{NFE} = 20$, consistent with the sound domain.

\begin{table*}[htbp]
\centering
\caption{Raw experimental results across three domains.}
\resizebox{\textwidth}{!}{%
\begin{tabular}{llcccccccccccc}
\toprule
\multicolumn{2}{c}{\textbf{Configuration}} & \multicolumn{6}{c}{\textbf{Speech Domain}} & \multicolumn{3}{c}{\textbf{Sound Domain}} & \multicolumn{3}{c}{\textbf{Music Domain}} \\
\cmidrule(lr){1-2} \cmidrule(lr){3-8} \cmidrule(lr){9-11} \cmidrule(lr){12-14}
\textbf{Method} & \textbf{BR} & \textbf{ZH-SIM} & \textbf{ZH-MOS} & \textbf{ZH-CER} & \textbf{EN-SIM} & \textbf{EN-MOS} & \textbf{EN-WER} & \textbf{FAD} & \textbf{KL} & \textbf{IS} & \textbf{FAD} & \textbf{KL} & \textbf{AES} \\
\midrule
\multirow{4}{*}{Large} 
& 0.975 & 0.885 & 4.073 & 0.075 & 0.821 & 3.969 & 0.041 & 96.93 & 0.463 & 7.069 & 130.78 & 0.103 & 7.067 \\
& 0.575 & 0.849 & 4.130 & 0.096 & 0.774 & 4.043 & 0.048 & 99.89 & 0.506 & 6.980 & 138.63 & 0.119 & 7.065 \\
& 0.375 & 0.810 & 4.153 & 0.096 & 0.724 & 4.082 & 0.058 & 106.38 & 0.558 & 6.910 & 148.18 & 0.126 & 7.060 \\
& 0.275 & 0.777 & 4.184 & 0.102 & 0.684 & 4.101 & 0.065 & 115.66 & 0.626 & 6.719 & 158.49 & 0.146 & 7.054 \\
\midrule
\multirow{4}{*}{Medium} 
& 0.975 & 0.826 & 4.170 & 0.094 & 0.756 & 3.969 & 0.056 & 126.85 & 0.664 & 6.774 & 179.10 & 0.167 & 7.101 \\
& 0.575 & 0.798 & 4.202 & 0.101 & 0.721 & 4.018 & 0.061 & 125.11 & 0.697 & 6.686 & 181.42 & 0.175 & 7.096 \\
& 0.375 & 0.773 & 4.215 & 0.109 & 0.685 & 4.060 & 0.066 & 127.57 & 0.740 & 6.667 & 182.46 & 0.183 & 7.076 \\
& 0.275 & 0.754 & 4.207 & 0.120 & 0.661 & 4.062 & 0.068 & 135.67 & 0.804 & 6.362 & 191.04 & 0.209 & 7.064 \\
\midrule
\multirow{4}{*}{Small} 
& 0.975 & 0.793 & 4.132 & 0.105 & 0.730 & 3.914 & 0.059 & 144.20 & 0.774 & 6.459 & 219.40 & 0.210 & 7.040 \\
& 0.575 & 0.769 & 4.166 & 0.109 & 0.697 & 3.970 & 0.064 & 142.58 & 0.766 & 6.311 & 220.40 & 0.221 & 7.030 \\
& 0.375 & 0.749 & 4.161 & 0.114 & 0.673 & 3.985 & 0.064 & 145.70 & 0.846 & 6.277 & 218.86 & 0.224 & 7.021 \\
& 0.275 & 0.734 & 4.151 & 0.119 & 0.651 & 4.007 & 0.069 & 150.48 & 0.887 & 6.268 & 220.15 & 0.228 & 7.016 \\
\bottomrule
\end{tabular}%
}

\label{tab:raw_metrics}
\end{table*}

\begin{table*}[t]
\centering
\footnotesize
\caption{Linear probing on codec letents across sound, music, and speech tasks.}
\label{tab:benchmark}
\resizebox{\textwidth}{!}{
\begin{tabular}{l cccc cccc cccc}
\toprule
\multirow{2}{*}{\bf Method} &
\multicolumn{4}{c}{\bf Sound} &
\multicolumn{4}{c}{\bf Music} &
\multicolumn{4}{c}{\bf Speech} \\
\cmidrule(lr){2-5}
\cmidrule(lr){6-9}
\cmidrule(lr){10-13}
&
{\bf ESC-50} & {\bf US8K} &{\bf FSD50K} & {\bf VIVAE} &
{\bf FMA} & {\bf MTT} & {\bf IRMAS} & {\bf MS-DB} &
{\bf RAVDESS} & {\bf AM} & {\bf SLURP} & {\bf EMOVO} \\
\midrule
WavTokenizer &16.45&29.24&4.02&28.48&30.50&16.99&23.19&47.66&28.82&56.70&7.21&17.69\\
UniCodec &24.20&37.55&5.21&27.58&34.88&21.97&25.16&53.70&26.04&64.67&7.44&22.11 \\
\midrule
Ours & {\bf 88.40}&{\bf 76.95}&{\bf 48.21}&{\bf 39.50}&{\bf 70.88}&{\bf 47.07}&53.72&{\bf 77.88}&{\bf 64.58}&{\bf 98.77}&{\bf 57.44}&{\bf 41.84}  \\
\quad-- w/o LM &73.90&72.43&36.81&33.69&66.75&45.70&{\bf 54.32}&67.45&54.17&72.41&20.35&33.67 \\
\bottomrule
\end{tabular}
}
\end{table*}

\section{Detailed Experimental Results of Figure~\ref{fig:tradeoff}}
\label{appendix:raw}

Table \ref{tab:raw_metrics} presents the complete evaluation results across \emph{Speech}, \emph{Audio}, and \emph{Music} domains for different model sizes and bandwidth settings. The visualization in Figure~\ref{fig:tradeoff} illustrates the trade-off between bandwidth and model performance. To construct the figure from the raw data in Table \ref{tab:raw_metrics}, the following processing steps were applied:

\begin{enumerate}
    \item \textbf{Normalization:} To eliminate differences in scale and unit across heterogeneous evaluation metrics and enable direct comparison and aggregation, all metrics were normalized to the range $[0,1]$ using min-max scaling:
    \begin{equation}
        x' = \frac{x - \min(x)}{\max(x) - \min(x)}
    \end{equation}
    
    \item \textbf{Metric Inversion:} For metrics where lower values indicate better performance (``Lower is Better''), specifically CER, WER, FAD, and KL divergence, the normalized scores were inverted so that $1.0$ represents the best performance:
    \begin{equation}
        x_{\text{final}} = 1 - x'
    \end{equation}
    
    \item \textbf{Domain Scoring:} Finally, a composite score for each domain was calculated by averaging the processed metrics: \textbf{Speech Score} (average of processed ZH-SIM, ZH-MOS, ZH-CER, EN-SIM, EN-MOS, and EN-WER), \textbf{Sound Score} (average of processed FAD, KL, and IS), and \textbf{Music Score} (average of processed FAD, KL, and AES).
\end{enumerate}

\section{Linear Probing on Codec Latent}
To verify that Eq.~\eqref{eq:lm} learns faithful semantic representations and facilitates downstream understanding, we evaluated linear probing on the ARCH benchmark~\cite{ARCH} and conducted an ablation on LM supervision in Table~\ref{tab:benchmark}. First, the semantic embeddings learned by the proposed method consistently outperform the purely acoustic baseline on the ARCH benchmark, indicating that language-model supervision effectively injects high-level semantic information into the latent space rather than merely preserving acoustic characteristics. Second, removing LM supervision (w/o LM) leads to substantial degradation in both probing accuracy and reconstruction quality across sound, music, and speech. These results demonstrate the critical role of language-model-based semantic grounding in shaping informative latent representations, which not only improve semantic understanding but also provide more effective guidance for audio reconstruction.


\section{Model Complexity Analysis}
\begin{table}[t]
\centering
\caption{Model complexity comparison.}
\label{tab:complexity}
\resizebox{\columnwidth}{!}{
\begin{tabular}{lccc}
\toprule
{\bf Method} & {\bf \# Params} & {\bf FLOPs} & {\bf RTF} \\
\midrule
SemantiCodec (NFE=50) & 921.72M & 0.59T &0.346  \\
WavTokenizer & 80.91M & 3.77G & <0.001 \\
UniCodec & 210.87M & 17.63G & <0.001 \\
Moss-Audio-Tokenizer & 1.6B & 91.30G & 0.007 \\
\midrule
Ours (NFE=10) & 1.8B & 0.85T & 0.042  \\
Ours (NFE=20) & 1.8B & 1.51T & 0.076 \\
\bottomrule
\end{tabular}
}
\vspace{-1mm}

\end{table}

Table~\ref{tab:complexity} presents the model complexity comparison. All real-time factors (RTFs) are measured on a single H100 GPU. Compared with baseline codecs, our method introduces a higher computational cost due to model size and the iterative ODE sampling process, and the complexity scales with the number of denoising function evaluations (NFEs). Nevertheless, the measured RTF remains below 0.1 even with 20 denoising steps, indicating practical inference efficiency on modern accelerators. More importantly, as demonstrated by the objective and subjective evaluations, our method consistently achieves substantially better reconstruction quality in the ultra-low-bitrate regime. These results suggest that the additional computational overhead is justified by the significant quality gains obtained under extremely low bitrate constraints, where preserving perceptual fidelity is particularly challenging.

\section{Visualization Analysis}
\label{append:visual}
\begin{figure*}
    \centering
    \includegraphics[width=1.0\linewidth]{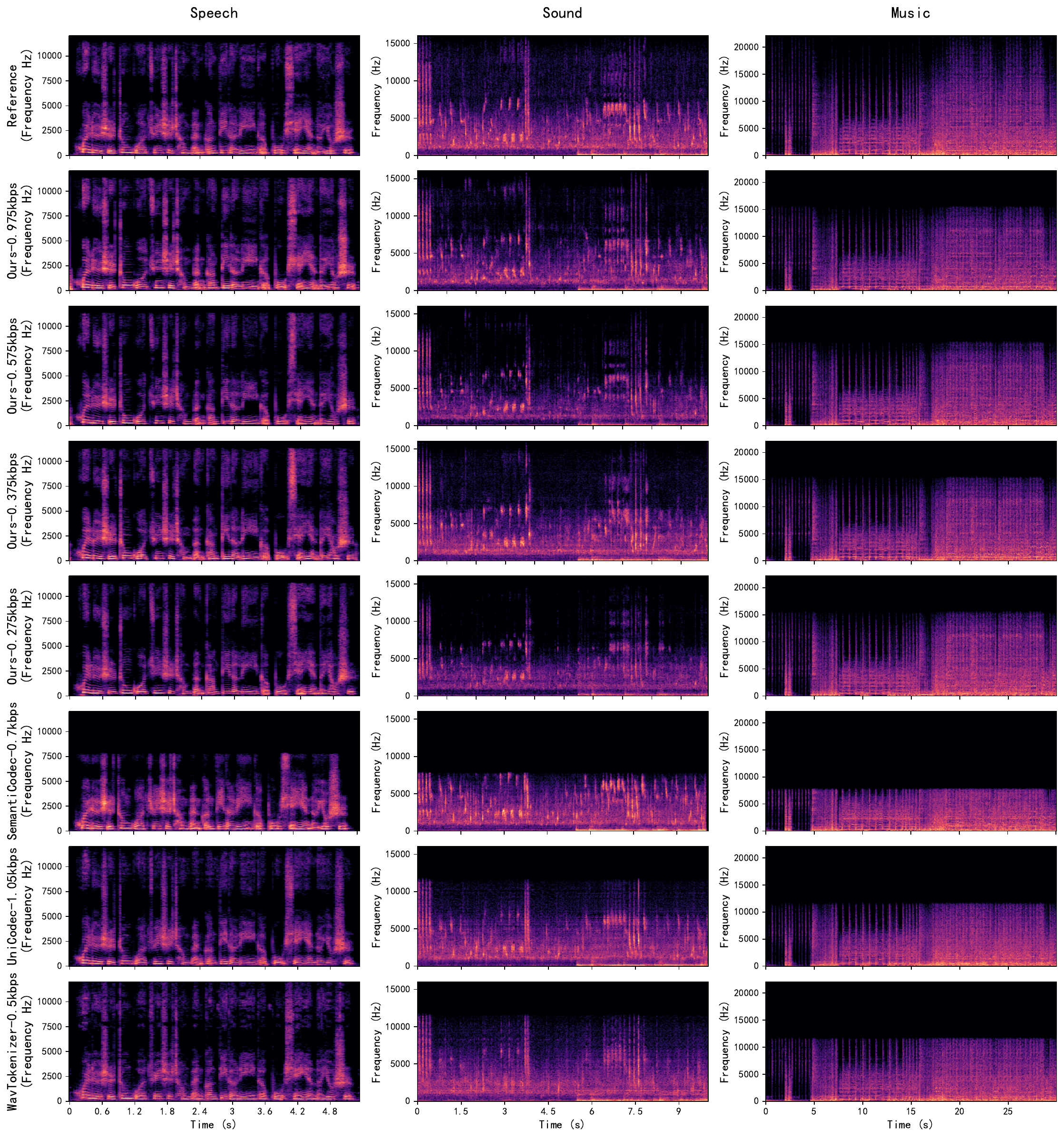}
    \caption{Spectrogram visualization of reconstructed audio samples from different audio codecs.}
    \Description{Spectrogram visualization of reconstructed audio samples from different audio codecs.}
    \label{fig:specvis}
\end{figure*}

To provide an intuitive comparison of reconstruction quality, we visualize the spectrograms of reconstructed audio samples from the speech, sound, and music domains. For each domain, we select one representative sample and compare the reconstruction results of our method with those of SemanticCodec~\cite{semanti}, UniCodec~\cite{unicodec}, and WavTokenizer~\cite{ji2024wavtokenizer}. As shown in Figure~\ref{fig:specvis}, our method better preserves the overall spectral structure and harmonic patterns, especially under ultra-low bitrate conditions. Compared with the baseline codecs, our reconstructions exhibit fewer spectral distortions and clearer temporal-frequency details, indicating improved semantic consistency and acoustic fidelity across different audio domains.

\end{document}